\documentclass[lettersize,journal]{IEEEtran}
\usepackage{newtxtext,newtxmath}
\usepackage{algorithmic}
\usepackage{algorithm}

\usepackage{array}
\usepackage[caption=false,font=footnotesize,labelfont=rm,textfont=rm]{subfig}
\usepackage{textcomp}
\usepackage{stfloats}
\usepackage{url}
\usepackage{verbatim}
\usepackage{graphicx}
\usepackage{cite}
\usepackage{booktabs}
\usepackage{multirow}
\usepackage{bm}
\usepackage{threeparttable}
\usepackage{tabularx}
\usepackage{colortbl}
\usepackage{color, xcolor} 
\usepackage{soul} 
\usepackage{balance} 
\definecolor{mygray}{gray}{.93}
\definecolor{mygreen}{RGB}{57, 124, 124}
\usepackage{xcolor}

\usepackage{hyperref}
\hypersetup{
	colorlinks=true,
	linkcolor=blue,
	filecolor=magenta,      
	urlcolor=cyan,
	pdftitle={Overleaf Example},
	pdfpagemode=FullScreen,
}

\begin{document}

\title{DS-TDNN: Dual-stream Time-delay Neural Network with Global-aware Filter for Speaker Verification}

\author{Yangfu Li, Jiapan Gan, Xiaodan Lin$^*$ and \\ School of Information Science and Engineering, Huaqiao University
\thanks{
* Corresponding Author. 
	
Yangfu Li, Jiapan Gan, and Xiaodan Lin are with the School of Information Science
and Engineering, Huaqiao University, Xiamen 361021, China (e-mail:
21013082029@stu.hqu.edu.cn; 22013082022@stu.hqu.edu.cn; xd\_lin@hqu.edu.cn).}
}

\markboth{Journal of \LaTeX\ Class Files, Jul~2023}%
{Shell \MakeLowercase{\textit{et al.}}: A Sample Article Using IEEEtran.cls for IEEE Journals}


\maketitle
\begin{abstract}
Conventional time-delay neural networks (TDNNs) struggle to handle long-range context, their ability to represent speaker information is therefore limited in long utterances. Existing solutions either depend on increasing model complexity or try to balance between local features and global context to address this issue. To effectively leverage the long-term dependencies of audio signals and constrain model complexity, we introduce a novel module called Global-aware Filter layer (GF layer) in this work, which employs a set of learnable transform-domain filters between a 1D discrete Fourier transform and its inverse transform to capture global context. Additionally, we develop a dynamic filtering strategy and a sparse regularization method to enhance the performance of the GF layer and prevent overfitting. Based on the GF layer, we present a dual-stream TDNN architecture called DS-TDNN for automatic speaker verification (ASV), which utilizes two unique branches to extract both local and global features in parallel and employs an efficient strategy to fuse different-scale information. Experiments on the Voxceleb and SITW databases demonstrate that the DS-TDNN achieves a relative improvement of 10\% together with a relative decline of 20\% in computational cost over the ECAPA-TDNN in speaker verification task. This improvement will become more evident as the utterance's duration grows. Furthermore, the DS-TDNN also beats popular deep residual models and attention-based systems on utterances of arbitrary length.
%

\end{abstract}


\section{Introduction}
\IEEEPARstart{A}{utomatic} Speaker Verification (ASV) systems that aims to determine whether a given utterance is from an enrolled speaker has been widely applied in user authentication, access control, multimedia forensics, and many others \cite{rosenberg1976automatic,broun2002automatic,becker2008forensic}. Typically, an ASV system consists of two main components: a front-end that extracts low-dimensional discriminative speaker embeddings from variable-length utterances and a back-end that determines whether two embeddings are from the same speaker. 
With the development of deep neural networks (DNNs), the front-end has shifted from probabilistic models \cite{reynolds2000speaker,kenny2007joint,dehak2010front} to DNN-based methods \cite{lei2014novel,variani2014deep,snyder2017deep}. In particular, x-vector \cite{snyder2017deep} as a state-of-the-art architecture for speaker embedding is built on the Time-Delay Neural Network (TDNN) \cite{waibel1989phoneme} layers for end-to-end speech processing, where shared-weight filters are employed to attend to all the frequency components and a time-delay strategy is applied to capture context between consecutive frames.

To extract robust speaker representation, both local features and global context are essential. However, typical TDNNs focus primarily on local features while being limited in modeling global context due to the small receptive field in each hidden layer.
As a result, TDNN-based models exhibit suboptimal performance in wild scenarios, particularly when the test utterances are exposed to noise. Furthermore, in real-world applications such as call monitoring, video processing, and real-time online meetings, utterances can last for tens of seconds, highlighting the need of global context modeling and faster inference speed for speaker verification.  A natural idea to emphasize global context is to extend the depth of TDNNs. \cite{novoselov2018deep} introduces residual connections \cite{he2016deep} to construct an extremely deep TDNN that captures context information over a longer time span. Snyder et al. \cite{snyder2019speaker} extend TDNN by inserting dense layers between each pair of hidden layers. To prevent overfitting, \cite{huang2019deeper} introduces dropouts \cite{srivastava2014dropout} to TDNN and proposes a filter with varying temporal resolution for more powerful context representation. To further deepen TDNN within affordable parameter overhead, Yu et al. \cite{yu2020densely} attempts to make a trade-off between the width and depth. However, thin TDNNs may suffer from less robust speaker representations. To solve this problem, a split-transform-merge structure is proposed in \cite{zhang2020aret} that helps the deep TDNN with limited width learn more discriminative speaker representations. 
These improvements make remarkable progress in performance, while they also suffer from the problem of model complexity, leading to poor real-time performance. Moreover, these methods do not consider the fusion of local and global features. 

\begin{figure}[!t]
	\centering
	\includegraphics[width=\linewidth]{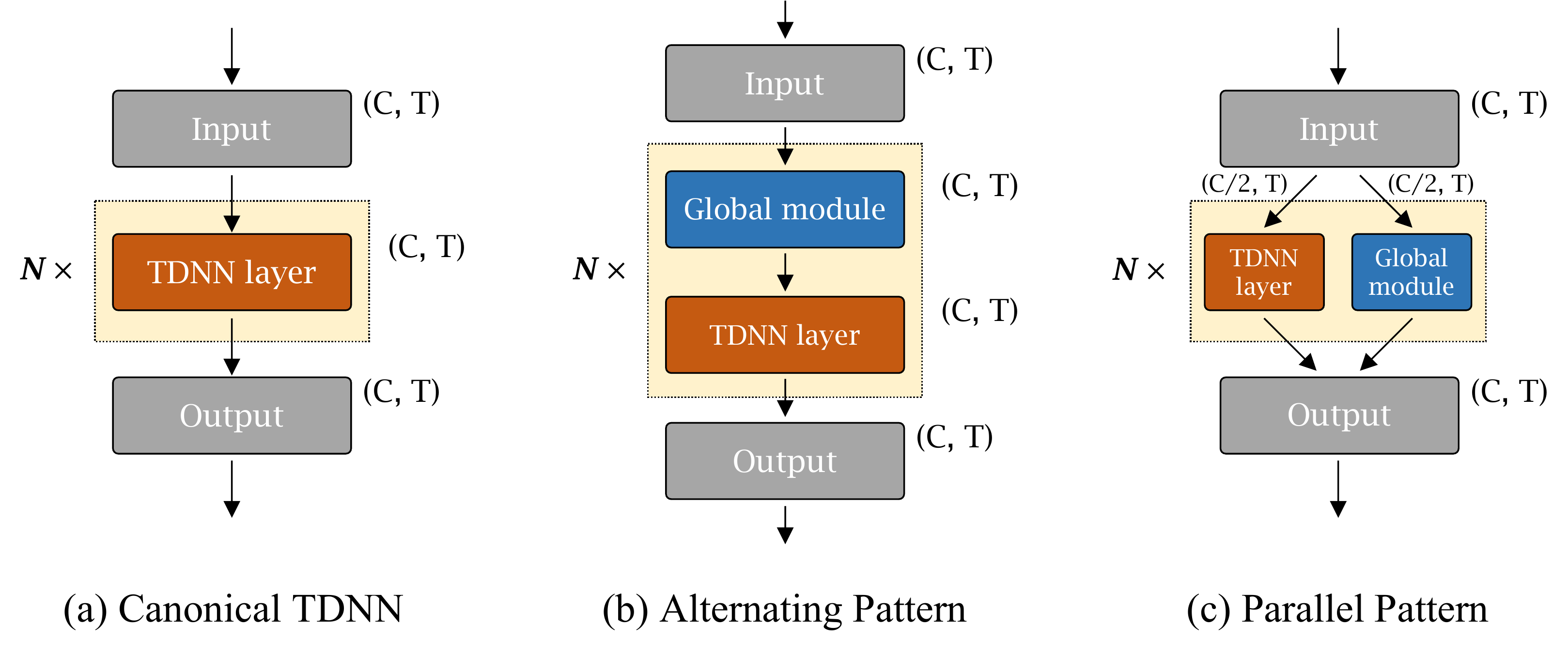}
	\caption{A comparison of different strategies to combine local and global features. (a) TDNN without additional modules for global context. (b) the popular alternating pattern. (c) the proposed parallel pattern.}
	\label{fig 1}
\end{figure}

To address the above issues, recent line of research can be roughly divided into two categories. A number of works is towards enhancing the filters with multi-scale information. \cite{desplanques2020ecapa} introduces the Res2Net \cite{gao2019res2net} structure into TDNN, by inserting skip connections between each grouped filter to construct a different-scale receptive field and proposes the multi-layer feature aggregation to fuse local and global features, achieving state-of-the-art performance. To enhance the Res2Net, \cite{9688119} proposes a context-aware filter by dividing each grouped filter applied in Res2Net into two different-scale filters: a normal receptive field filter for local features and a large receptive field filter for global context. Authors in \cite{gu2021dynamic} propose another type of dynamic filter, whose value is determined by an element-wise summation between the local features calculated using moving average and the global context calculated through global average pooling. Recently, \cite{mun2022selective} presents a multi-resolution filter, which employs a kernel selection mechanism to find the optimal receptive field. Although these enhanced filters can dynamically adapt to local features or global context as required, none of them simultaneously attend to both local features and global context, limiting their representation capability.
The other line of research considers incorporating the additional modules to handle global context with TDNN, e.g., autoregressive models and self-attention, as shown in Fig \ref{fig 1}. For example, authors in \cite{chen2019speaker} employ the Long Short-Term Memory (LSTM) module in the shallow layers of TDNN to provide high-resolution temporal context information. In \cite{jiang2019effective}, the authors insert LSTM between each pair of hidden layers in TDNN, enabling TDNN to capture global context. However, these dense LSTM methods are computationally expensive. To improve efficiency, \cite{sak2014long, liu2019speaker} utilize recurrent projection to reduce the dimension of the data flow in the LSTM cells. Another approach is to apply the LSTM at the segment-level rather than the frame-level, handling temporal information in the low-dimensional latent space \cite{tu2019towards,lin2019lstm}. With the success of self-attention, the authors in \cite{huang2020speaker} combine the frame-level LSTM with the segment-level attention mechanism to further enhance the performance. This technique has shown promising performance, but it comes with the side effect of high computational cost.

The existing designs of the global modules, such as LSTM or self-attention, have intensive computational complexity. Additionally, different-scale operations are directly performed on the whole mixture of local and global features as illustrated in Fig. \ref{fig 1}b, incurring high parameter overhead. To address these two challenges, this paper first proposes a novel and efficient global module called the Global-aware Filter (GF) layer. The GF layer consists of three key components: a discrete Fourier transform, a set of differentiable transform-domain filters, and the inverse discrete Fourier transform. It explicitly models the global context, but only has a log-linear complexity. Moreover, we introduce dynamic filtering to further enhance the GF layer. With dynamic filtering, the GF layer can adapt to different speech contexts, thereby improving its representation ability and generalization performance. To prevent over-fitting, we also propose sparse regularization, which randomly drops filters in the GF layer with a fixed ratio during the training phase.
In addition, we propose a parallel workflow to combine local and global features, as shown in Fig. \ref{fig 1}c. In this pattern, different-scale modules are designed to focus on certain parts of the complete feature maps, which is more efficient than the popular alternating pattern. By incorporating the GF layer and the parallel framework, we construct a simple yet efficient dual-stream TDNN for speaker verification, called DS-TDNN. DS-TDNN employs two independent branches to process local features and global context in parallel and applies several carefully designed strategies to fuse different-scale information. Experimental results on the Voxceleb and SITW datasets demonstrate that DS-TDNN outperforms the powerful TDNN-based baseline, ECAPA-TDNN, with a lower computational cost. Furthermore, it outperforms other popular baseline systems, such as 2D CNN with residual connections and the attention-based model. Moreover, DS-TDNN achieves the best trade-off between effectiveness and efficiency. We have released the models and code for further research\footnote[1]{\url{https://github.com/YChenL/DS-TDNN}}. 

Our contributions are summarized as follows:
\begin{itemize} 
	\item We propose an innovative module termed \emph{Global-aware Filter} (GF) layer for TDNN, which has global receptive fields yet a log-linear complexity wrt. speech duration. The GF layer is an efficient alternative to popular global-aware algorithms in speaker verification.

	\item We design two special techniques, i.e., \emph{Dynamic Filtering} and \emph{Sparse Regularization} to further enhance the performance of the GF layer. The former enables the GF layer dynamically adapt to the input, providing a more powerful representation of speaker information. The latter aims to reduce the optimization difficulty of the dynamic GF layer and prevent overfitting.
	
	\item We propose a novel parallel framework named DS-TDNN for the extraction of different-scale information, as a solution to speaker embedding. Experiments on the Voxceleb and SITW datasets demonstrate that the proposed DS-TDNN achieves state-of-the-art performance compared to popular baseline systems.
	

\end{itemize}

The rest of this paper is organized as follows: Section II introduces the global-aware filter layer, including its motivation, implementation, improvements, and comlexity analysis. Section III describes the dual-stream TDNN. The experimental setup is detailed in Section IV. The result analysis is presented in Section V, followed by a conclusion in Section VI.

\section{Global-aware filter}

\subsection{Motivation}
Conventional TDNNs mainly utilize time-delay layers for feature extraction, where convolutions are restricted within the window rather than the entire feature map, limiting the modeling for long-range context. 
A natural idea for expanding the receptive field of convolution is to increase the window size. For example, employing the global convolution\cite{8099672} to capture the global context. However, the global convolution results in a $\mathcal{O}(N^2)$ complexity for a sequence including $N$ points. Fortunately, there is a simple yet efficient equivalent of the global convolution, which can be derived from the discrete convolution theorem 
and is formulated as:
\begin{equation}
	\bm {w}_g* \bm x \equiv \mathcal{F}^{-1}[{\bm w}_f\mathcal{F}[\bm x]],
\end{equation}
where $\bm w_f$ is regarded as a transform-domain filter. $\bm w_g$ is a spatial-domain filter with a global receptive field, $\bm x$ is an $N$-point token.
$*$ denotes one-dimensional convolution. $\mathcal{F}[\,\bm \cdot\,]$ and $\mathcal{F}^{-1}[\,\bm \cdot\,]$ separately denote the one-dimensional discrete Fourier transform (DFT) and its inverse transform (IDFT). Benefitting from the fast algorithms of DFT/IDFT, i.e., FFT/IFFT, the complexity of $N$-point DFT/IDFT is reduced from $\mathcal{O}(N^2)$ to $\mathcal{O}(N{\rm log}N)$. Equ 1 reveals the use of FFT/IFFT to model long-range context, which has been demonstrated powerful in computer vision tasks\cite{li2020falcon,chi2020fast,lifourier}, but has never been explored for ASV. Motivated by this, we introduce the global-aware filter layer for TDNN that modulates the tokens using FFT/IFFT together with a set of differentable filters $\bm w_f$ to handle the global context. 

\subsection{Global-aware filter Design}
We propose a global-aware filter (GF) layer as an efficient alternative to global-aware modules, e.g., global convolution or self-attention, to capture global context. In TDNN, the tokens $\bm X \in \mathbb{R}^{C\times T}$ can be considered a series of discrete sequences $\bm x_i\in \mathbb{R}^{1\times T}$ stacked along the channel dimension. Therefore, given the tokens ${\bm X}$, the corresponding spectrum $\bm X_f$ can be obtained by 1D FFT individually performed on every channel, where the number of FFT bins is the same as the length of each input channel.
\begin{equation}
	{\bm X_f}={\rm Concat}(\mathcal{F}[{\bm x_1}], \mathcal{F}[{\bm x_2}], ..., \mathcal{F}[{\bm x_C}])\in \mathbb{C}^{C\times T}.
\end{equation}
For efficiency, we perform channel-wise 1D FFT/IFFT on $\bm X$ and then employ linear projection to mix channels, rather than directly performing 2D FFT/IFFT on $\bm X$.
Besides, since $\bm X$ is a real tensor, the corresponding spectrum $\bm X_f$ is conjugate symmetric, i.e., ${\bm X_f}[: , T-\tau]={\bm X_f^*}[: , \tau]$. Thus, only half of $\bm X_f$ is needed for further processing:
\begin{equation}
	{\bm X_r}={\bm X_f}[:, 0:\lceil T/2\rceil]= \mathcal{F}_r[{\bm X}] \in \mathbb{C}^{C\times \lceil T/2\rceil},
\end{equation}
where $\mathcal{F}_r[\,{\bm \cdot}\,]$ denotes the channel-wise 1D FFT for real signals. 
It is noteworthy that $\bm X_r$ is a complex tensor. 
Since the Fourier transform integrates the whole information of a channel into different elements of the spetrum, we can model the channel-wise global context via a simple element-wise multiplication between the spectrum and a differentiable filter $\bm F \in \mathbb{C}^{C\times \lceil T/2\rceil}$:
\begin{equation}
	\tilde{{\bm X_r}} = {\bm F} \odot {\bm X_r},
\end{equation}
where $\odot$ is the Hadamard product. $\bm F$ can be regarded as the $\bm w_f$ in Eq.(1), termed global-aware filter. Finally, the inverse FFT is adopted to transform the modulated spectrum $\tilde{\bm X_r}$ back and update the tokens: 
\begin{equation}
	{\bm X} \leftarrow \mathcal{F}^{-1}_r[{\tilde{\bm X_r}}].
\end{equation}

Depending on the type of filter $\bm F$, the GF layer can selectively capture either local features or global context of the tokens. When a high-pass filter is used, the GF layer tends to capture local features, while a low-pass filter allows the GF layer to capture global context.

The proposed GF layer can easily adapt to different audio lengths as both the FFT and the IFFT have no learnable parameters. For different lengths of utterances, we can simply interpolate the global-aware filter $\bm F$ to $\bm F' \in \mathbb{C}^{C\times \lceil T^{\prime}/2\rceil} $  where $T^{\prime}$ is the target length. From the frequency sampling point of view,  the process of duration adaptation is equivalent to resampling of the spectrum. Besides, since FFT/IFFT are well supported by GPU and CPU, the GF layer is hardware friendly. 

\subsection{Dynamic filtering}
The challenge of ASV in wild speech comes from the intervention of speech content, emotion, and transmisssion channel. Therefore, the distribution between utterances from the same speaker may vary significantly, making it hard to generalize with the static filters. To address this issue, we propose a dynamic filtering strategy inspired by \emph{Dynamic Convolution} that enables the filters to adapt dynamically to the input \cite{yang2019condconv,chen2020dynamic,li2022omni}. Specifically, we apply $K$ independent global-aware filters during modulation and combine them using element-wise summation:

\begin{equation}
	\tilde{{\bm X_r}}= {\bm F_1}\odot {\bm X_r} + {\bm F_2}\odot {\bm X_r}+...+{\bm F_K}\odot {\bm X_r}.\\
\end{equation}
Then, we utilize a 1D channel attention function to produce a series of dynamic scores $\bm w=[w_1, w_2,...,w_K]$, which can be adapted to the input tokens $\bm X$: 
\begin{equation}
	{\bm w} = \underbrace{{\tt Softmax(FC_2(ReLU(FC_1(GAP}}_{\rm Attention\ FN}({\bm X}))))) \in\mathbb{R}^{1\times K},
\end{equation}
where ${\tt GAP}$ is Global Average Pooling. ${\tt FC_1}$ and ${\tt FC_2}$ separately represent two fully connected layers. In this work, the number of neurons in $\tt FC_1$ is equal to that in $\tt FC_2$, which is set to $K$. Notably, we utilize Softmax rather than Sigmoid to normalize the scores for stable training. The dynamic scores $\bm w$ are then adopted to weight the expert filters, which is formulated as:
\begin{equation}
	\tilde{{\bm X_r}}=w_1{\bm F_1}\odot {\bm X_r}+ w_2{\bm F_2}\odot {\bm X_r}+...+ w_K{\bm F_K}\odot {\bm X_r}.\\
\end{equation}
For simplicity, an equivalent deformation is obtained as
\begin{equation}
	\tilde{{\bm X_r}}=(w_1{\bm F_1}+ w_2{\bm F_2}+...+ w_K{\bm F_K})\odot {\bm X_r},\\
\end{equation}
from which the \emph{dynamic global filter} (DGF) is defined to represent the normalized linear combination of filters as
\begin{equation}
	{\bm F_d} = w_1{\bm F_1}+ w_2{\bm F_2}+...+w_K{\bm F_K}\in \mathbb{C}^{K\times C\times\lceil T/2\rceil}.
\end{equation}
Finally, the spectrum $\bm X_r$ is modulated by the element-product with the dynamic global-aware filter $\bm F_d$.

\subsection{Sparse regularization}

Despite that dynamic filtering is helpful to rendering a more robust and generalized representation, it also poses a challenge regarding the optimization procedure, i.e., a total of $K$ global dynamic filters are to be optimized instead of a static one, resulting in a more complicated loss landscape. Additionally, as the number of parameters increases, the model becomes prone to overfitting. To address this issue, we propose a sparse regularization technique. Specifically, the DGF can be viewed as a stack of 1D filters ${\bm F_d}=[\bm f_1, \bm f_2, …, \bm f_d]$, where $\bm f_1,…,\bm f_d \in \mathbb{C}^{1\times \lceil T/2\rceil}$. During the training phase, parts of these filters are deactivated via an element-wise multiplication with a random sparse-channel mask $\bm M$, i.e., ${\bm F_s}=\bm M\odot {\bm F_d}$. Hereafter ${\bm F_s}$ is defined as the dynamic global-aware filter with sparse regularization (sparse DGF). Notably, to maintain the dynamic characteristic, the mask is directly applied to the DGF rather than to each of the expert filters. During the modulation with sparse DGF, we perform an element-wise summation to preserve the information at regions where the deactivated filters are applied.

\begin{equation}
	\tilde{\bm X}_r = {\bm F_s}\odot{\bm X_r}+\lambda_s(1-\bm M)\odot{\bm X_r},
\end{equation}
where $\lambda_s=\frac{1}{CT}\sum^C_{i=1}\sum^T_{j=1}|\bm F_{d}^{(i,j)}|$ is a scale factor. 
The unactivated filter can be regarded as a scaled all-pass filter. Noteworthily, the sparse regularization is conducted only in the training phase.

\begin{figure*}[!t]
	\centering
	\includegraphics[width=\linewidth]{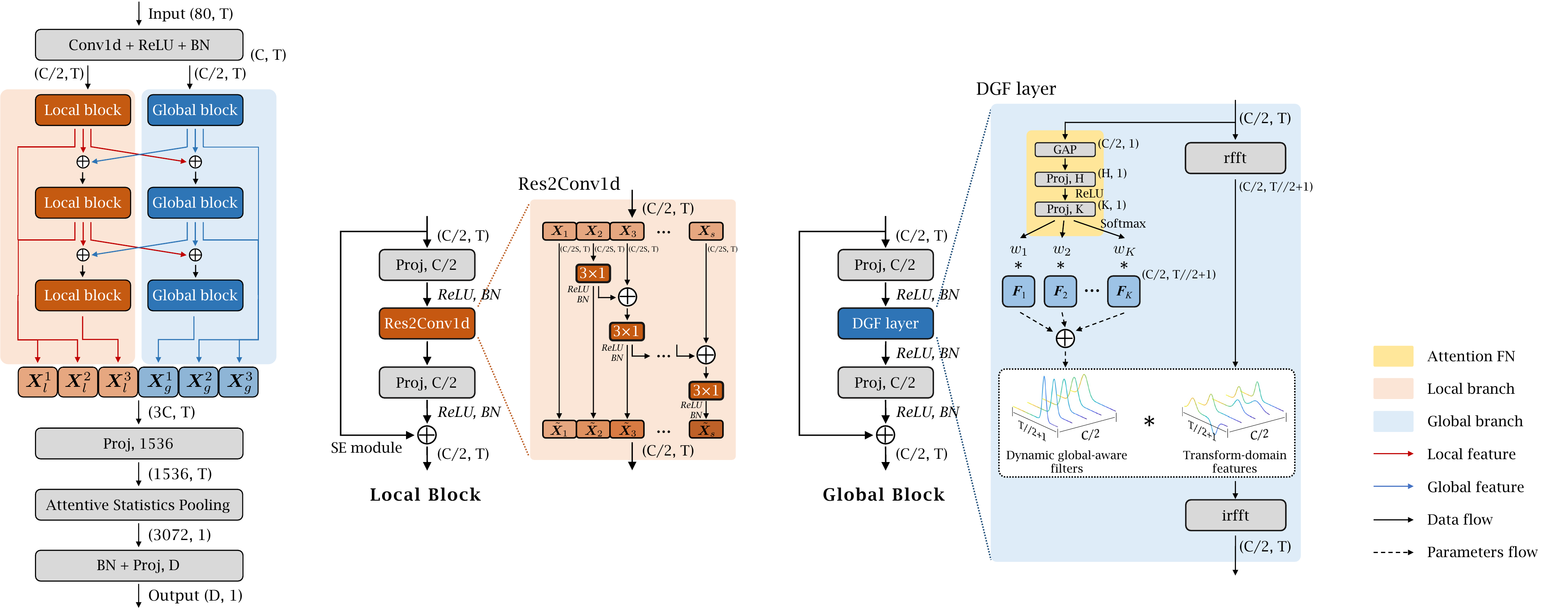}
	\caption{Overview of the DS-TDNN architecture, where $\oplus$ denotes element-wise summation; $*$ denotes element-wise multiplication; BN denotes 1D BatchNorm; (Proj, C) is a linear projection layer with C hidden neurons; GAP is global average pooling; 3$\times$1 is a 1D convolution with kernel size of 3 and step of 1. }
	\label{fig 2}
\end{figure*}

\section{Dual-stream TDNN}

In addition to efficiently handling long-range context, it is also crucial to consider how to combine local and global features in the model. Existing works perform different-scale operations alternately over the entire feature maps that consist of both local and global features, as shown in Fig. \ref{fig 1}b. However, this alternating pattern has two shortcomings. Firstly, applying local filters to extract global context or global filters to extract local features is inefficient. Secondly, to create a more discriminative speaker representation, feature maps in TDNN generally feature a large size, making it computationally expensive. To address this issue, we propose a parallel framework, as shown in Fig. \ref{fig 1}c. Different modules are designed to pay attention to complementary features rather than the entire inputs, and the output of each module is concatenated as the final result. We assume that the input feature maps contain equal number of local features and global context. Under this assumption, the channel dimension of the dataflow in the parallel framework is reduced to half compared to the alternating pattern. Notably, only a simple channel split is required for the disentanglement of the local features and the global context. Based on this idea, we propose the dual-stream (DS-TDNN) model, which applies the DGF layer (detailed in Section II) as the global module and combines the local features and global context in the parallel pattern.

\subsection{Macro design}
The architecture of DS-TDNN is shown in Fig. \ref{fig 2}. Overall, DS-TDNN is comprised of two inter-connected branches: a local branch that integrates the Res2Conv module to capture local features, and a global branch that modulates spectrums using the DGF layer in transform domain to handle long-range context. Each branch operates on only half of the input channels in parallel and captures complementary information with different-scale receptive fields. The input of each branch is extracted from 80-dimension Mel spectrogram $\bm X \in \mathbb{R}^{80\times T}$, then passes through a 1D convolution with a kernel size of 7 and a step of 1, ReLU activation, and 1D BatchNorm:
\begin{equation}
	{\bm X_{l}^0}, {\bm X_{g}^0} = {\tt Split}({\tt BN}({\tt ReLU}({\tt Conv}({\bm X })))) \in\mathbb{R}^{C/2\times T},
\end{equation}
where ${\bm X_{l}^0}$, ${\bm X_{g}^0}$, denote the inputs of the local and global branches, respectively. $C$ represent the numbers of basic channels in the DS-TDNN. 
Besides the final fusion of the two braches, normalized element-wise summation is also employed to gradually fuse local and global features within the branch. The output of the $i$-th layer in the branch can be formulated as follows:
\begin{equation}
	\begin{aligned}
		&{\bm X_{l}^i}={\tt Block}_{l}^i(0.8{\bm X_{l}^{i-1}}+0.2{\bm X_{g}^{i-1}}) \in \mathbb{R}^{C/2\times T},\\
		&{\bm X_{g}^i}={\tt Block}_{g}^i(0.2{\bm X_{l}^{i-1}}+0.8{\bm X_{g}^{i-1}}) \in \mathbb{R}^{C/2\times T}, 
	\end{aligned}
\end{equation}
where $1\leq i\leq N$. ${\bm X_{l}^i}$, ${\bm X_{g}^i}$ denote the output of the $i$-th local and global block. ${\tt Block}_l^i$, ${\tt Block}_g^i$ represent the $i$-th local block and global block respectively. Then, a multi-scale feature aggreation (MFA) \cite{desplanques2020ecapa,liu2022mfa,zhang2022mfa} is performed to fuse different-scale speaker information by concatnating the output of each layer. Following this, a linear projection is employed to fuse the different-scale  information as
\begin{equation}
{\bm H}={\tt Proj(Concat}({\bm X^1_l},{\bm X^2_l},{\bm X^3_l},{\bm X^1_g},{\bm X^2_g},{\bm X^3_g}))\in \mathbb{R}^{\hat{C}\times T},
\end{equation}
where the projection dimension $\hat{C}$ is set to 1536 in this work. Then, the Attentive Statistics Pooling (ASP) \cite{okabe2018attentive} is applied to weight the importance of each frame-level feature $\bm h_t \in \mathbb{R}^{\hat{C}\times 1}$ of $\bm H$ and extract the robust speaker embedding, which is given by 
\begin{align}
	e_t &= {\bm v}^T{\tt Tanh}({\bm W}{\bm h_t} +{\bm b})+k,\\
    \alpha_t &= \frac{{\tt exp}(e_t)}{\sum^T_{\tau=1}{\tt exp}(e_\tau)},
\end{align}
where $\bm W\in\mathbb{R}^{\hat{C}\times \hat{C}}, \bm v\in \mathbb{R}^{\hat{C}\times 1}, \bm b\in \mathbb{R}^{\hat{C}\times 1}$ and $k$ are the learnable parameters for ASP. After that, the normalized score $\alpha_t$ is adopted to calculate the weighted mean vector $\tilde{\bm \mu}$ and weighted standard deviation $\tilde{\bm \sigma}$, yielding
\begin{align}
	\tilde{\bm \mu}&=\sum^T_{t=1}\alpha_t {\bm h_t},\\
	\tilde{\bm \sigma}&=\sqrt{\sum^T_{t=1}\alpha_t {\bm h_t}\odot {\bm h_t}-\bm \mu\odot\bm \mu},
\end{align}
where  $\bm \mu=\frac{1}{T}\sum^T_{\tau=1}{\bm h_\tau}$. The output of the ASP is given by concatenating the vectors of the weighted mean $\tilde{\bm \mu}$ and weighted standard deviation $\tilde{\bm\sigma}$. Finally, the speaker embedding is reduced to a low dimensional vector with 1D BatchNorm and linear projection.

\begin{table*}[t]
	\begin{center}
		\caption{Configurations of the Variants of DS-TDNN and the ECAPA-TDNN Peers.}
		\label{Table 1}
		\renewcommand\arraystretch{1.5}
		\setlength{\tabcolsep}{3mm}{
			\begin{tabular}{l|cccccc}
				\bottomrule
				Model     &Blocks [Local, Global] &Channels $C$  &Scales $s$  &Experts $K$ &Sparse ratio &\#Params(M)\\
				\hline
				\hline
				ECAPA-c512 & $[1,\ 0]\times 3$ & $[512,\ 0]$      & $[8,\ 8,\ 8]$  &-   &-           & 7.0\\
				DS-TDNN-S  & $[1,\ 1]\times 3$ & $[256,\ 256]$      & $[4,\ 4,\ 4]$  & $[4,\ 4,\ 8]$ & $[0.3,\ 0.1,\ 0.1]$ 
				& 6.5\\
				\hline
				\hline
				ECAPA-c1024& $[1,\ 0]\times 3$ & $[1024,\ 0]$     & $[8,\ 8,\ 8]$  &-   &-           & 15.5\\
				DS-TDNN-B  & $[1,\ 1]\times 3$ & $[512,\ 512]$     & $[4,\ 4,\ 8]$  & $[4,\ 8,\ 8]$ & $[0.3,\ 0.1,\ 0.1]$ 
				& 13.2  \\	
				\hline
				\hline
				ECAPA-L   & $[1,\ 0]\times 3$ & $[1280,\ 0]$     & $[8,\ 8,\ 8]$  &- & -          & 21.1 \\
				DS-TDNN-L  & $[1,\ 1]\times 3$ & $[768,\ 768]$   & $[4,\ 8,\ 8]$  & $[8,\ 8,\ 8]$ & $[0.4,\ 0.2,\ 0.2]$
				& 20.5 \\
				\toprule
		\end{tabular}}
	\end{center}
\end{table*}

\subsection{Micro design}
Each branch consists of three macaron-like blocks. In the blocks, two linear projections sandwich the filter applied for token mixing. The first projection is utilized to disentangle the local (global) information from the mixture, while the second projection is employed to exchange the channel information.
\subsubsection{Local block} 
To enhance the capability for local feature representation, \emph{Res2Conv} structure \cite{gao2019res2net,desplanques2020ecapa} is applied in local blocks as the token mixer. As shown in Fig. \ref{fig 2}, the Res2Conv is a combination of group convolutions. It firstly splits the input tokens $\bm X_l^i \in \mathbb{R}^{C/2\times T}$ into $s$ groups in channel dimensions:
\begin{equation}
	{\bm X_l^i} = [{\bm X_1}, {\bm X_2},...,{\bm X_s}],
\end{equation}
where ${\bm X_1}, {\bm X_2},...,{\bm X_s}\in\mathbb{R}^{C/2s\times T}$ represent the token groups. Subsequently, a 1D convolution with a kernel size of 3 and a step of 1 is applied in each group, followed by ReLU activation and 1D BatchNorm. Notably, since the local block is designed to focus on local features, the dilation rate of the convolutions is set to 1 in this work. Besides, convolution is not applied to the first group of tokens to lower the computation cost. The input of each convolution layer is the sum of the token group and the output of the previous convolution layer:
\begin{equation}
	\tilde{{\bm X}}_{i+1} = {\tt ReLU(\tt BN(\tt Conv}({\bm X_{i+1}}+\tilde{\bm X}_{i}))), i=2,...,s,
\end{equation}
where $\tilde{{\bm X}}_{i+1}$ represents the output of the $(i+1)$-th convolution layer. The output of the Res2Conv1d is a concatenation of $\tilde{{\bm X}}_{i+1}, i=1,...,s$, which provides different-scale features extracted from ${\bm X_l^i}$. Finally, the SE module \cite{hu2018squeeze} is applied at the end of each local block.

\subsubsection{Global block} 
As shown in Fig. \ref{fig 2}, the structure of global blocks is similar to that of local blocks, where a DGF layer is utilized to replace Res2Conv to capture long-range context information from long-time span. In addition, we perform a simple skip connection rather than the channel attention module used in local blocks, since the Attention FN shown in Eq.(7) has already contained discriminative information about channels.

\subsection{Architecture variants}
To evaluate our model, we developed three variants of the DS-TDNN. The first two variants, i.e., DS-TDNN-S and DS-TDNN-B, have similar hyperparameters as those of the typical ECAPA introduced in \cite{desplanques2020ecapa}. Additionally, we investigated variants of ECAPA and DS-TDNN of a larger size, i.e., ECAPA-L and DS-TDNN-L, that have similar parameters to enable a fair comparison between TDNN-based models and other baseline systems, such as transformer-based models and deep residual 2D CNNs. Table \ref{Table 1} provides a summary of the detailed configurations of these variants. As shown in Table \ref{Table 1}, due to the efficient DGF layer and the proposed parallel framework, the parameters of DS-TDNN are less than the ECAPA counterpart, which becomes more evident as the number of channels increases. By default, we applied sparse regularization to all variants of the DS-TDNN.

\section{Experimental Setup}

\subsection{Data Preparation}

VoxCeleb1 \& 2 \cite{nagrani2017voxceleb, chung2018voxceleb2}, and SITW \cite{mclaren2016speakers} are used in our experiments. VoxCeleb is an audio-visual dataset consisting of over 2,000 hours of short clips of human speech extracted from interview videos on YouTube. SITW is a widely-used standard evaluation dataset collected from open-source media in real-world conditions, and is made up of 299 speakers, including two testing trials (SITW.Dev and SITW.Eval) that have over 2,800 utterances from 180 speakers. All the systems are trained only on the development set of VoxCeleb2, which has over 1,092,009 utterances at a sampling rate of 16 kHz from 5,994 speakers. A small subset of about 2\% of the data is reserved as a validation set for hyperparameter optimization.

To better illustrate the advantages of global context modeling for utterances of different duration, we conducted four trials: VoxCeleb1-O (i.e., Vox1-O), VoxCeleb1-E (i.e., Vox1-E), VoxCeleb1-H (i.e., Vox1-H), and a mixture consisting of SITW.Dev and SITW.Eval (i.e., mix-SITW) for performance evaluation. Specifically, VoxCeleb1-O is the test part of VoxCeleb1, which contains 40 speakers with a total of 37,720 test pairs sampled from VoxCeleb1. VoxCeleb1-E is an extension of VoxCeleb1-O, including 1,251 speakers with a total of 581,480 test pairs. VoxCeleb1-H is the more challenging scenario, including 552,536 test pairs where the country and gender of the speakers in each pair are the same. Most of the utterances in VoxCeleb1 last for 5-8 seconds, thus Vox1-O, Vox1-E, and Vox1-H can be regarded as short-duration utterances. As for SITW, the major durations are about 30–40 seconds. Therefore, the mix-SITW is adopted to simulate the long-duration scenario. 
To further investigate how the performance changes as the duration of utterances increases, we randomly clip the test utterance of the mix-SITW with a step size of 5s, thus yielding four duration settings: $\leq 5$s, $\leq 15$s, $\leq 30$s, and $\leq 50$s, containing a total of 2500 test pairs.
Notably, the training set used in the experiments, i.e., the development set of VoxCeleb2, is completely disjoint from these four evaluation trials.

As data augmentation is generally effective for improving the performance of neural networks, we apply six augmentation strategies following the Kaldi recipe \cite{snyder2019speaker} in combination with the publicly available MUSAN dataset \{music, speech, noise\} \cite{snyder2015musan} and the RIR dataset \{reverberation\} \cite{ko2017study}. Each of the five datasets contributes equally to the augmented training dataset in an additive way, i.e.,  reverberation, speech, music, noise, and a mixture of speech and music are added to the speech corpus. The last augmentation strategy applying to all of the training samples is SpecAugment \cite{park2019specaugment}, which randomly masks 0 to 5 frames and 0 to 10 channels of the log Mel spectrogram.

\subsection{System description}

In order to comprehensively evaluate the performance of the proposed DS-TDNN, not only the TDNN-based models, i.e., \emph{ECAPA-TDNN} \cite{desplanques2020ecapa}, but also the 2D CNN-based models, i.e., \emph{SE-ResNet} \cite{chung2020defence,zhao2021speakin,shim2022graph}, and transformer-based models, i.e., \emph{Audio Spectrogram Transformer (AST)} \cite{gong2021ast} and \emph{MFA-Conformer} \cite{zhang2022mfa}, are regarded as baseline systems. The inputs for all systems are 2-second Mel spectrograms of 80 dimensions generated from a 25ms window with a 10ms frame shift, and the speaker embedding dimension is 192. For fair comparison, we tune their hyper-parameters slightly. The configurations of the baseline systems are introduced as follows:

\textbf{AST}: It is a fully attention-based model, taking the Mel spectrogram as input and producing speaker embeddings using the ASP on the averaged tokens produced by a transformer encoder. The dimension of its input is reduced from 128 to 80 in our experiments to make it consistent with other baseline systems. Two variants of AST are applied in our experiments, i.e., AST-tiny (AST-T) and AST-small (AST-S). The AST-T consists of 12 self-attention layers with 3 heads and 192 hidden channels, while the AST-S has 12 self-attention layers with 6 heads and 384 hidden channels.  

\textbf{MFA-Conformer}: It is a hierarchical attention-based model, which introduces convolution to provide the local information, achieving the state-of-the-art performance for speaker verification. It has 6 macaron-like blocks with 1/2 subsampling rates, where two feed-forward networks (FFN) sandwich the composition of multi-head self-attention (MSA) and convolution. In the following, the FFNs have 2048 hidden units, the MSA has 4 heads with 272 dimensions, and the convolution has a kernel size of 15 and a step of 1. Besides, the ASP is employed before producing the final speaker embeddings.

\textbf{SE-ResNet}: SE-ResNet have similar structure and hyper-parameters to ResNet while applying channel attention (SE module) to improve speaker verification performance. Notably, the number of basic channels of SE-ResNet is set to 32 rather than 64 in our experiments. Besides, the subsampling rate of the stem is set to 1/2 instead of the commonly used 1/4 for all variants. In addition, ASP is used in all variants to replace the statistics pooling in our experiments.

\textbf{ECAPA-TDNN} \& \textbf{DS-TDNN}: Three pairs of variants, i.e., ECAPA-c512 and DS-TDNN-S, ECAPA-c1024 and DS-TDNN-B, ECAPA-L and DS-TDNN-L, are investigated as peer work based on TDNN , whose settings are detailed in Table \ref{Table 1}.

\subsection{Training strategy}
To minimize the duration mismatch between the training and the evaluation, we employed a two-stage training process. The first stage involved pre-training for 150 epochs, followed by large margin fine-tuning (LM-FT) \cite{thienpondt2021idlab} for 5 epochs. During pre-training ($\sim 35$ hours), all systems are trained using Additive Angular Margin (AAM) loss \cite{deng2019arcface, xiang2019margin}, with the margin and scale set to 0.2 and 30.0, respectively. Adam \cite{kingma2014adam} is used as the optimizer, with an exponentially decreasing learning rate from $10^{-3}$ to $10^{-6}$. To avoid overfitting, we set the weight decay to $2\times10^{-5}$ and perform a linear warmup for the first 2k steps. During LM-FT ($\sim 1.2$ hours), we increase the duration of training samples to 6 seconds and raise the margin to 0.5. The learning rate is initialized to $10^{-4}$ and decreased to $2.5\times10^{-5}$, with a batch size of 512. All the experiments are conducted on 4$\times$NVIDIA RTX A5000.


\subsection{Backend}
Speaker embeddings are extracted from the final fully connected layer for all systems. Trial scores are produced using the cosine distance between embeddings. Subsequently, adaptive score normalization (as-norm) \cite{cumani2011comparison} is used to normalize the trial score. We average the embeddings from the same speaker in the training set to construct the imposter cohort and set the imposter cohort size to 600. Performance is measured by the equal error rate (EER) and the minimum normalized detection cost (minDCF) with $P_{target} = 0.01$ and $C_{FA} = C_{Miss} = 1$. In addition, the real-time factor (RTF) calculated by the Intel Xeon Platinum 8358P (2.60GHz) is also provided to evaluate the inference speed of different models.

\begin{table*}[!t]
	\caption{Voxceleb EER (\%) and minDCF comparison among different models. ‘-c’ denotes the number of basic channels. The best results are marked in \textbf{BLOD}, the second are marked \underline{UNDERLINE}. Performance of our models are highlighted in \protect\sethlcolor{mygray}\hl{GRAY}.}
	\centering
	\label{Table 2}
	\renewcommand\arraystretch{1.3}
	\setlength{\tabcolsep}{2mm}{
		\begin{tabular}{ll |ccc cc cc cc }
			\bottomrule
			\multirow{2}{*}{Index} & \multirow{2}{*}{Backbone} & \multirow{2}{*}{FLOPs (G)}& \multirow{2}{*}{\#Param (M)} & \multirow{2}{*}{RTF ($\downarrow$)}&  \multicolumn{2}{c}{Vox1-O} & \multicolumn{2}{c}{Vox1-E} & \multicolumn{2}{c}{Vox1-H}\\
			\cline{6-11}
			& & &  & & EER (\%) & minDCF & {EER (\%)} & minDCF & EER (\%) & minDCF \\
			\hline		
			\hline	
			N1  & AST-T 
			&1.1 &7.0  & 0.0071  &1.61  &0.170  &1.98  &0.208  &3.42  &0.296         \\
			N2  & SE-ResNet34      
			&1.2 &6.4  & \underline{0.0053}  &1.15  &0.149  &1.41  &0.166  &2.75  &0.253         \\		
			N3  & ECAPA-c512   
			&1.2 &7.0  & \textbf{0.0046}  &\underline{1.04}&\underline{0.133} &\underline{1.26} &\underline{0.151} &\underline{2.36} &\underline{0.224} \\
			\rowcolor{mygray}
			N4  & DS-TDNN-S    
			&1.0 &6.5  & 0.0058   &\textbf{0.90}  &\textbf{0.118} &\textbf{1.15} &\textbf{0.140} &\textbf{2.11} &\textbf{0.199}\\ 
			
			\hline
			\hline 
			N5  & SE-ResNet50  
			&1.4 &11.9 & 0.0092  &1.05          &0.124          &1.25          &0.156          &2.17          &0.206         \\	  
			N6  & SE-ResNet101   
			&2.6 &17.0 & 0.0149  &0.90          &\underline{0.107} &1.14 &0.143  &\underline{1.94}         &\underline{0.186}\\ 
			N7  & ECAPA-c1024  
			&2.9 &15.5 & \underline{0.0069}  &\underline{0.88} &0.114       &\underline{1.12}&\underline{0.135}&2.08      &0.202         \\	\rowcolor{mygray}
			N8  & DS-TDNN-B  
			&2.1 &13.2 &  \textbf{0.0066}    &\textbf{0.78} &\textbf{0.092} &\textbf{1.06} &\textbf{0.126} &\textbf{1.86} &\textbf{0.174}\\	
			\hline
			\hline
			N9  & AST-S
			&4.4  &22.5 &  0.0134  &1.08  &0.125  &1.40     &0.152  &2.38     &0.216  \\  
			N10 & MFA-Conformer  
			&2.1  &20.8 & \textbf{0.0071}  &\underline{0.70} &\underline{0.087} &\underline{0.99} &\underline{0.120} &\underline{1.64}&\underline{0.158}\\
			N11 & SE-ResNet152
			&3.8  &21.8 & 0.0258    &0.75  &0.094  &1.09     &0.128  &1.82     &0.176  \\	
			N12 & SE-ResNet34-c64 
			&4.7  &23.6 & 0.0132    &0.98  &0.122  &1.21     &0.147  &2.13     &0.196  \\		
			N13 &ECAPA-L   
			& 4.0 & 21.1& 0.0088  & 0.79 & 0.106 & 1.08 & 0.131 & 1.87 & 0.181\\	
			\rowcolor{mygray}	
			N14 & DS-TDNN-L 
			&3.2  &20.5 & \underline{0.0083}  &\textbf{0.64} &\textbf{0.082}   &\textbf{0.93} &\textbf{0.112} &\textbf{1.55} &\textbf{0.149} \\	
			\toprule
	\end{tabular}}
\end{table*}

\section{Results and Analysis}


\subsection{Results on Voxceleb}
The performance comparisons of DS-TDNN and various baseline systems introduced in Section IV.B are reported in Table \ref{Table 2}, which is measured by the equal error rate (EER) and minimum Detection Cost Function (minDCF) together with the number of model parameters, floating point operations (FLOPs), and real time factor (RTF). The FLOPs and RTF are measured on 5-second utterances.

The results of experiment N1–N4 demonstrate that the proposed DS-TDNN-S system achieves the best recognition performance among popular tiny baseline systems, although its RTF is not the best. For the midium-sized systems in N5–N8, DS-TDNN-B has fastest inference speed and the best recognition performance. Specifically, compared with the typical ECAPA-TDNN system with 1024 channels, the proposed DS-TDNN-B system achieves about an 11\% improvement in recognition performance with a 15\% decline in parameters. For larger systems in N9-N14, the proposed DS-TDNN-L also outperforms in all the evaluation metrics except that its inference speed is a bit inferior to MFA-Conformer. Three groups of experiments demonstrate that as the network goes deeper, the performance of ASV could be improved. Noteworthily, it's observed from N11 and N12 that depth extension is more effective than width extension for 2D CNNs, emphasizing the importance of a larger receptive field for the ASV task. Moreover, the results in N9-N10 demonstrate that incorporating convolution in the transformer-based model could significantly enhance the recognition performance, highlighting the importance of local features. Finally, for systems of any size, the proposed DS-TDNN architecture always beats its peer models, indicating the advantage of combining local features and global context for speaker verification.

\begin{figure*}[!t]
	\centering
	\subfloat[]{
		\includegraphics[width=0.30\linewidth]{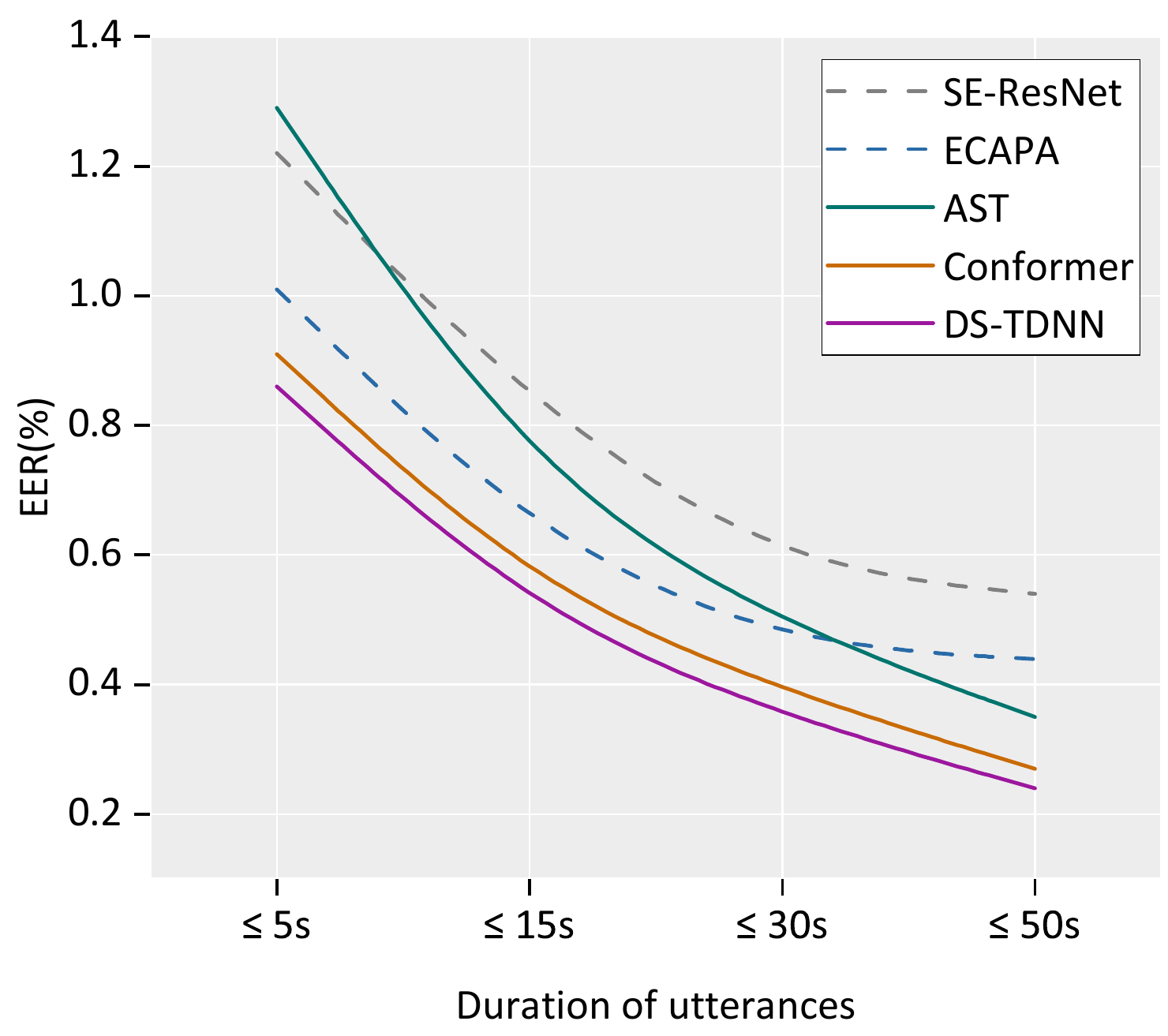}
	}
	\subfloat[]{
		\includegraphics[width=0.30\linewidth]{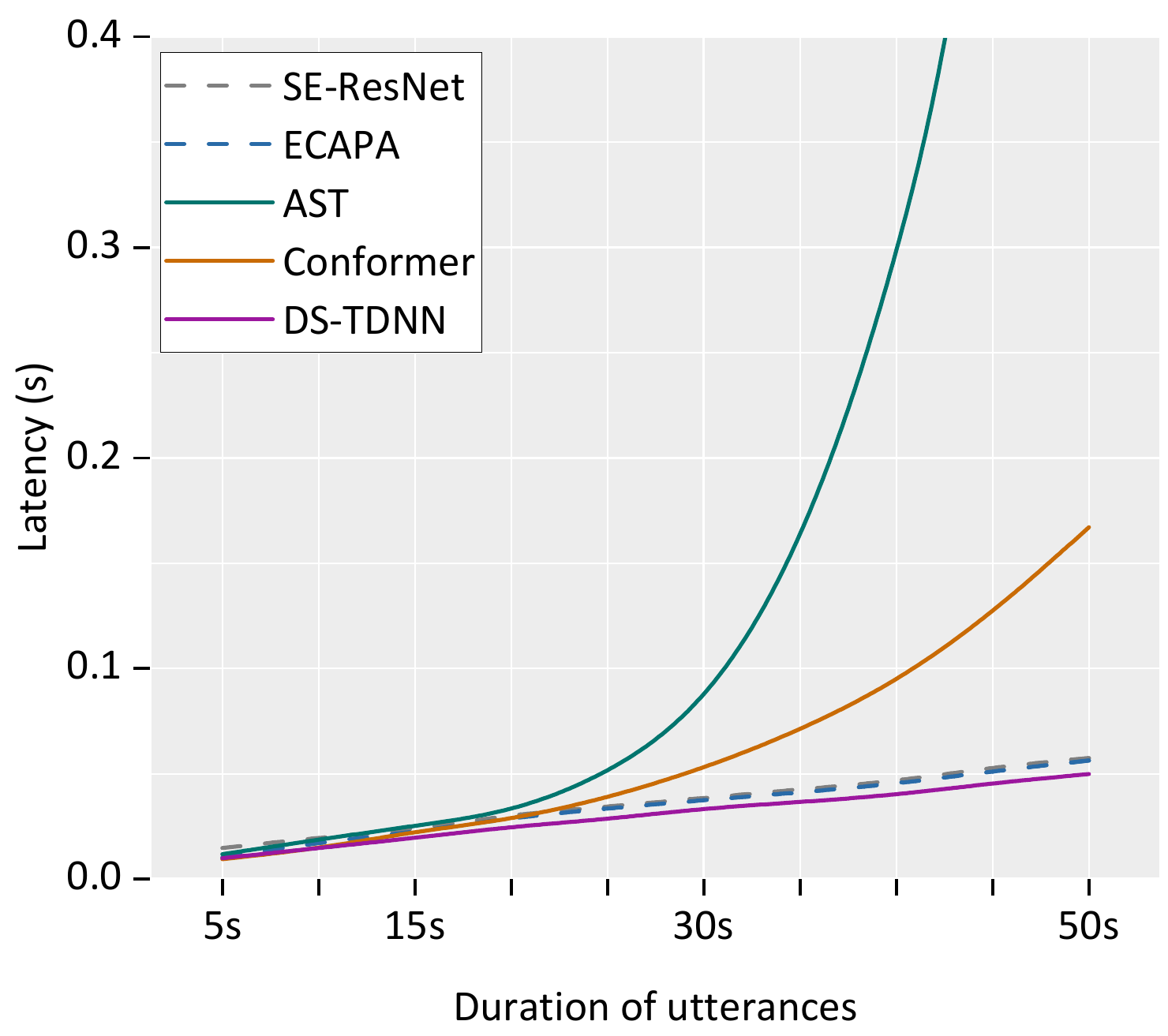}
	}
	\subfloat[]{
		\includegraphics[width=0.30\linewidth]{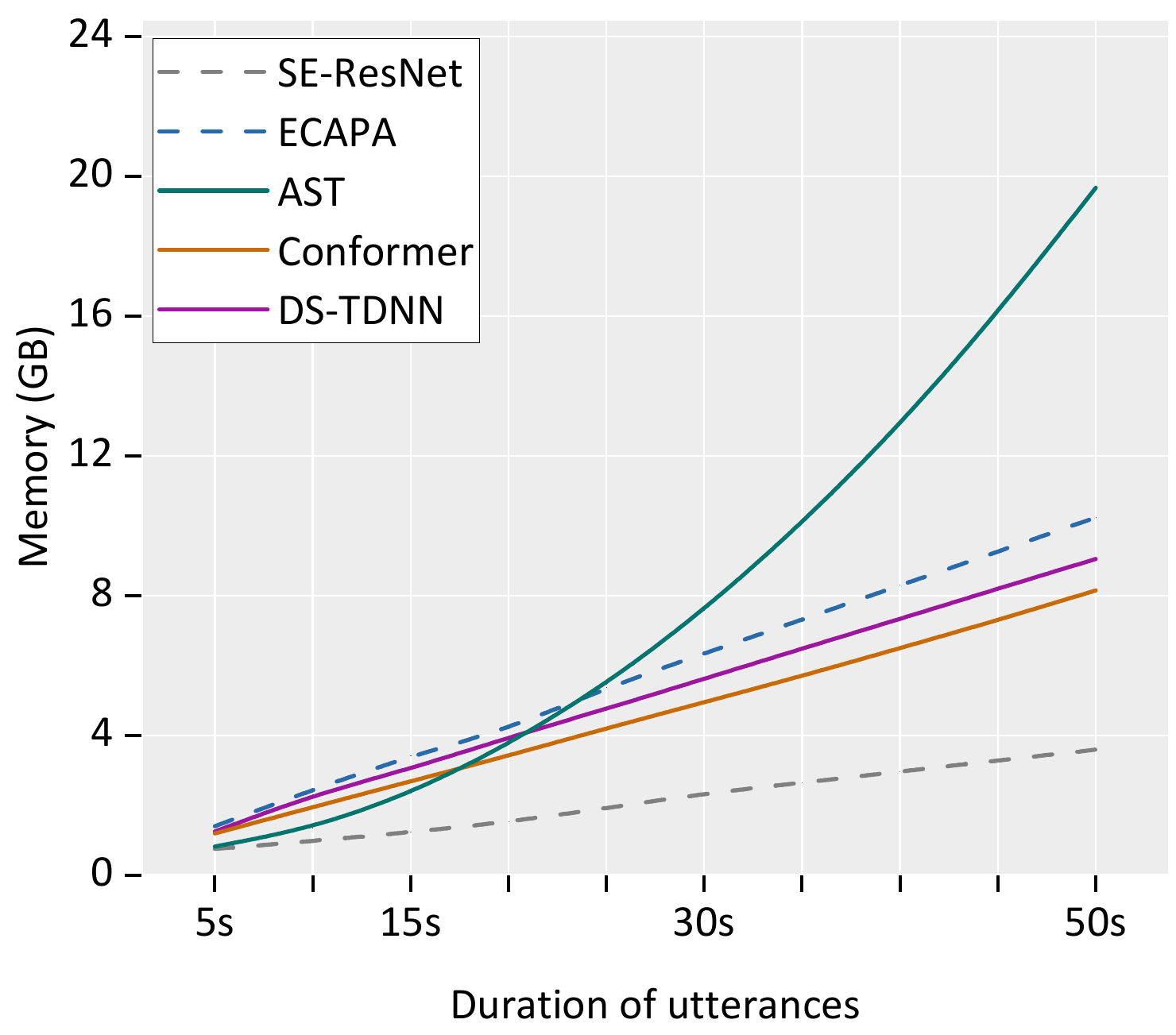}
	}
	\caption{Comparisons among SE-ResNet34-c64 \cite{chung2020defence}, ECAPA-TDNN-L\cite{desplanques2020ecapa}, AST-S \cite{gong2021ast}, MFA-Conformer \cite{zhang2022mfa} and the proposed DS-TDNN-L under different settings of utterance's duration. (a) EER results (b) Latency and (c) Memory usage. The dash lines correspond to models without the special design to handle long-range context. The latency and memory usage is measured using a single NVIDIA A5000 GPU with batch size 16.}
	\label{fig 3}
\end{figure*}

\subsection{Results on SITW}

To evaluate the impact of utterance's duration, we make a comprehensive assessment on the mix-SITW dataset, using five models in Table \ref{Table 2}, including SE-ResNet34-c64, ECAPA-L, AST-S, MFA-Conformer, and the proposed DS-TDNN-L. The EER result and computational overhead are presented in Fig. \ref{fig 3}.

\begin{figure*}[!t]
	\centering
	\hspace{-4mm}
	\includegraphics[width=0.90\linewidth]{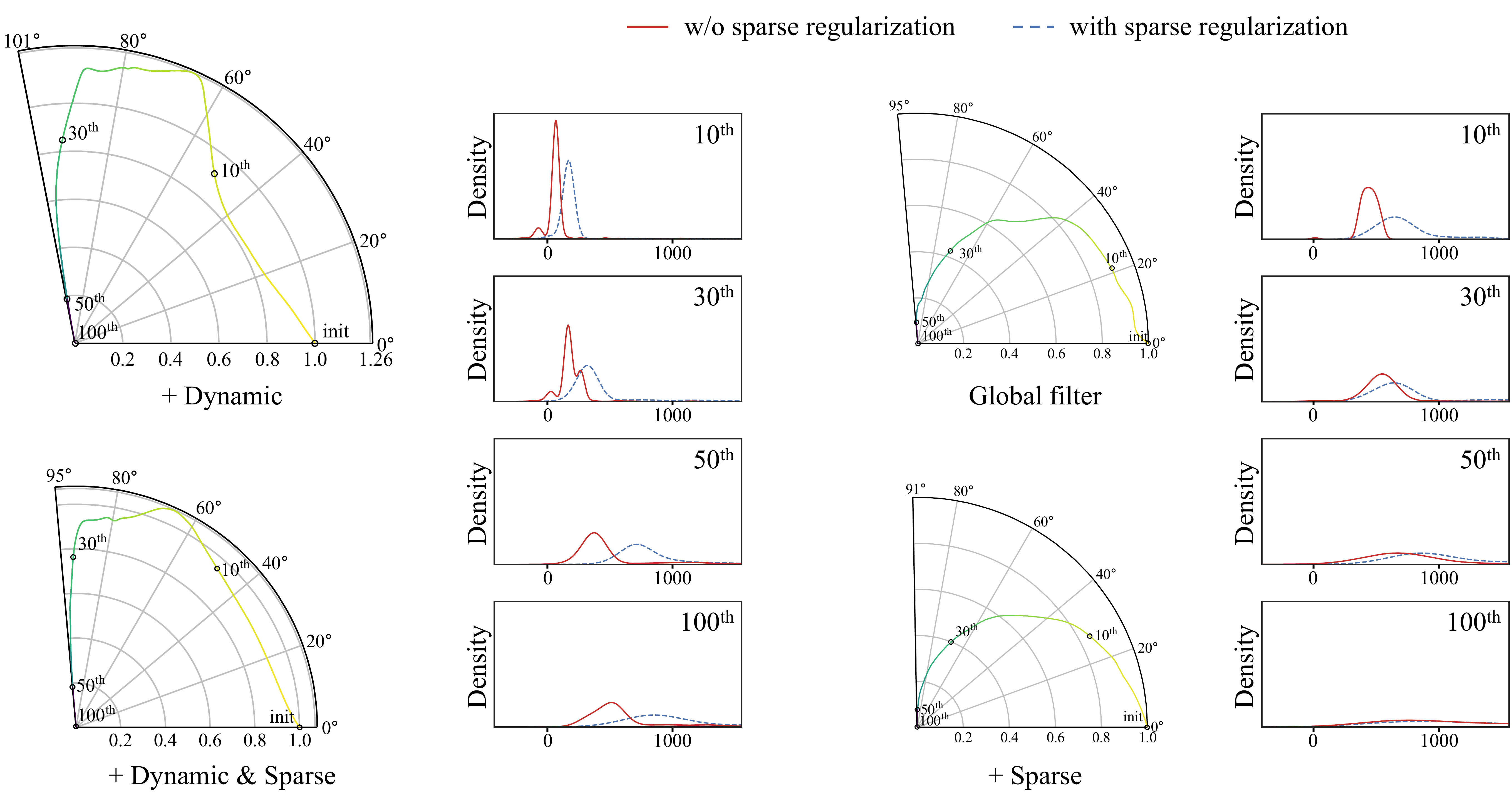}
	\caption{Visualizations of dynamic filtering and sparse regularization. \emph{Left}: DS-TDNN converges to the optimum along a smooth trajectory. \emph{Right}: The distribution of maximum Hessian eigenvalue. In the polar coordinate, $r_t=\frac{||\Delta\omega_{t}||}{||\Delta\omega_{\rm init}||}$, and $\theta={\rm cos^{-1}}\left(\frac{\Delta\omega_t\cdot\Delta\omega_{\rm init}}{||\Delta\omega_t||\,||\Delta\omega_{\rm init}||}\right)$, where $\Delta\omega_t=\omega_t-\omega_{\rm optim}$.}
	\label{fig 4}
\end{figure*}

Fig. \ref{fig 3}a demonstrates that DS-TDNN performs exceptionally well in ASV task and exhibits strong generalization capabilities in complex, real-world scenarios. Given speech signals of varying durations, DS-TDNN always outperforms the other four systems. Furthermore, the DS-TDNN, along with the AST and MFA-Conformer, consistently improves in performance as the duration increases, highlighting the importance of global context in extracting robust speaker representations from longer utterances. In contrast, SE-ResNet and ECAPA-TDNN models have limited receptive fields, making it hard to capture global context information from longer speech signals. As a result, the performance of these methods is surpassed by AST when the duration of test utterances exceeds about 30 seconds. Apparently, MFA-Conformer and DS-TDNN perform well for any audio length, since they attend to both local and global features. In terms of efficiency, DS-TDNN has the lowest inference latency on the GPU, attributed to the log-linear complexity of the DGF layer. Notably, the computational cost of the AST model increases significantly due to the quadratic complexity of self-attention with regard to input lengths. To address this issue, the MFA-Conformer employs a well-designed FFN to down-sample the feature map. However, its inference speed is still slower than that of the other three baseline systems when the duration of test utterances exceeds 25 seconds, as shown in Fig. \ref{fig 3}b. Additionally, Fig. \ref{fig 3}c shows that the memory usage of DS-TDNN is comparable to that of ECAPA-TDNN and MFA-Conformer, but higher than that of SE-ResNet and lower than that of AST. Overall, DS-TDNN strikes an impressive balance between speaker verification performance and computational cost, implying its promising application in real life.

\subsection{Analysis and visualization}

\subsubsection{Optimization analysis}

In order to investigate the effect of dynamic filtering and sparse regularization on the optimization procedure, the optimization trajectory in polar coordinates is plotted in Fig. \ref{fig 4} considering different filtering strategies, where the radius $r_t$ is defined as the normalized distance between the current weights $\omega_t$ and the optimum weights $\omega_{\rm optim}$, and the angle represents the direction of optimization. In addition, we also study the distribution of the Hessian eigenvalue of the weights in the dynamic/static GF layers, which reflects the local convexity of the loss function that indicates the training difficulty. 

From the top-left of Fig. \ref{fig 4}, the optimization trajectory of the model becomes significantly sharper after adopting dynamic filtering, and there is a detour in the optimization process, indicating that the dynamic filtering would increase the optimization difficulty. The detours diminish after taking sparse regularization as shown in the bottom-left of Fig. \ref{fig 4}. One possible explanation is that the sparse regularization reduces the variance of gradients for mini-batches and allows the model to update towards a consistent direction. This can also be verified by the distribution of the Hessian eigenvalue. Specifically, the Hessian of the loss has negative eigenvalues in early epochs, which indicates that the loss function is non-convex and the model is prone to falling into saddle points. Besides, although the Hessian eigenvalues yielded by DGF are smaller than those yielded by GF, the proposed sparse regularization significantly suppresses the negative eigenvalues, thus faciliating the training process. Interestingly, it is observed in the right part of Fig. \ref{fig 4} that sparse regularization also benefits GF.

\begin{figure*}[!t]
	\centering
	\includegraphics[width=\linewidth]{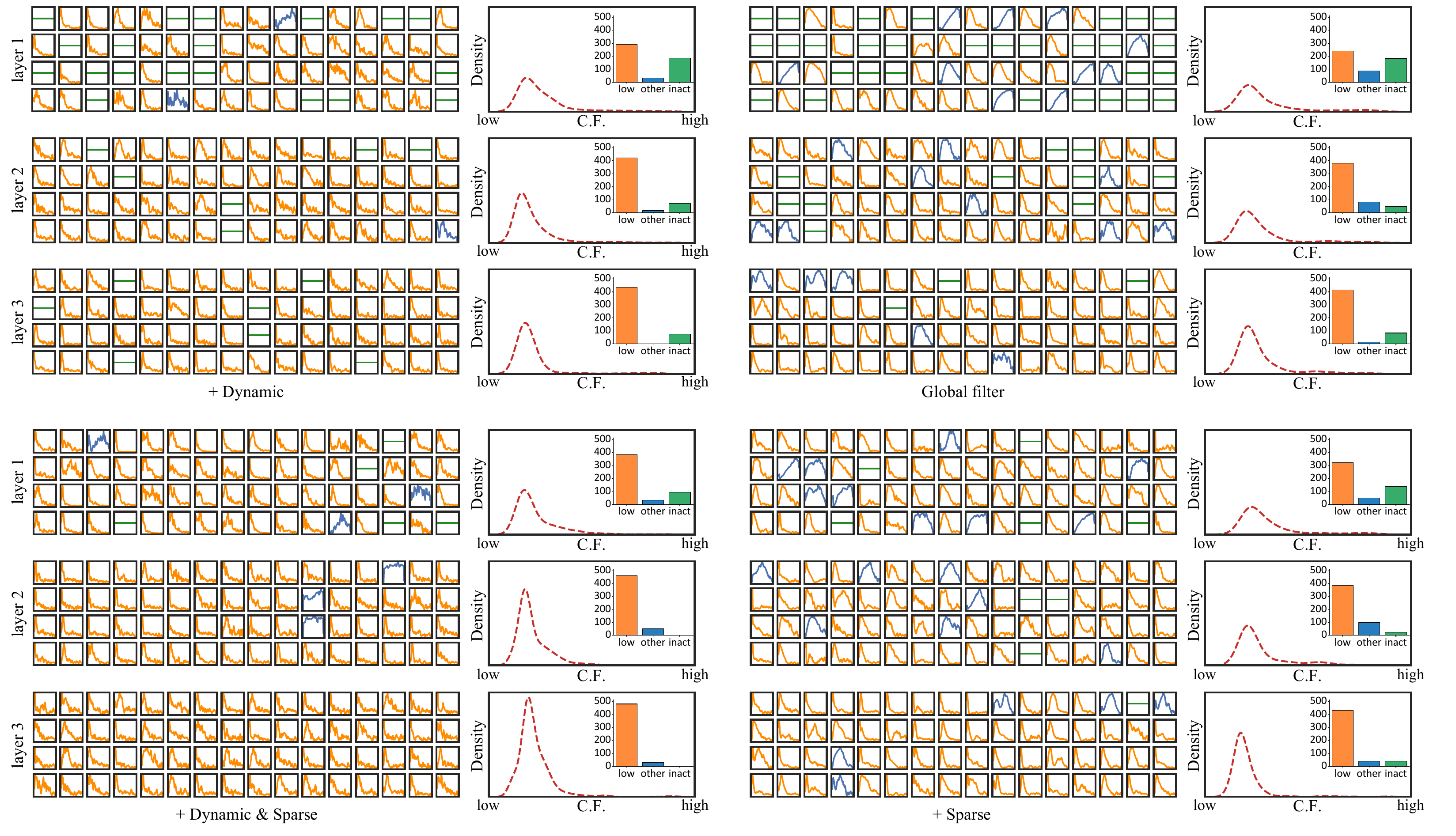}
	\caption{Visualizations of filters in the global-aware branch. \emph{Left}: the amplitude-frequency response of the first 64 filters in GF/DGF layers, where the vertical axis represents the log amplitude and the horizontal axis represents the frequency. \emph{Right}: the distribution of the center frequency (C.F.) of all the filters and the count of each type of filter.}
	\label{fig 5}
\end{figure*}

\subsubsection{Analysis of intermediate filters} 

In this part, we explore how the proposed GF affects input tokens. Besides, the effectiveness of dynamic filtering and sparse regularization for enhancing the GF is illustrated from the perspective of signal processing. For the DGF, we set $\omega_1 = \omega_2 = ... = \omega_K = 1/K$.

Firstly, numerous works have proven that the effect of global context modeling, e.g., self-attention, are equivalent to low-pass filters for the input tokens, while the local modeling like convolution tends to be high-pass filters \cite{parkvision,cordonnierrelationship,siinception}. As shown in Fig. \ref{fig 5}, both GF and DGF are mainly composed by low-pass filters, demonstrating that the long-term context rather than the local features is more emphasized. Therefore, DS-TDNN has the potential to capture more robust speaker representation from long-duration utterances than conventional TDNN. Secondly, from the comparison between GF and DGF, greater number of lowpass filters is found in DGF. More low-pass filters generally indicate stronger ability in global context modeling. Notably, there are some all-pass filters in the GF/DGF. Since these filters do not influence the spectrum, we call them inactive filters (inact). The existence of these inactive filters can be accounted by incomplete optimization. Nevertheless, it is easy to observe that the number of inactive filters is significantly reduced after sparse regularization, and the overall center frequency of GF/DGF is shifted to the lower region, implying that sparse regularization is helpful in the optimization process. Finally, the last two layers have more low-pass filters than the first layer, which is consistent with the fact that deeper layers of neural networks have larger receptive fields.

\section{Ablation study}

To determine the contribution of each component in the DS-TDNN, we conduct a detailed ablation study divided into four parts. First, we compare the DGF layer with three typical global-aware designs, i.e., multi-head self-attention (MSA), long short-term memory (LSTM), and bidirectional LSTM (Bi-LSTM). Second, we first investigate the macro designs of DS-TDNN, i.e., the combination of local features and global context, element-wise summation for different-scale information exchange, and multi-scale feature aggregation (MFA). Then, we evaluate the effects of dynamic filtering and Res2Conv on verification performance, respectively. Third, we quantitatively assess the impact of sparse regularization on dynamic global-aware filters and static filters. Finally, we explore the scalability of DS-TDNN from its depth and width. Besides, we compare the proposed parallel pattern with the conventional alternating pattern for feature combinations. Notably, we only present the EER and minDCF results on VoxCeleb1-O, as shown in Tables \ref{Table 5}-\ref{Table 8}, while similar trend is observed in other datasets.

\begin{table}[t]
	\caption{Ablation study on the global module design. }
	\centering
	\label{Table 5}
	\renewcommand\arraystretch{1.5}
	\setlength{\tabcolsep}{1.2mm}{
		\begin{tabular}{l| l l l l}
			\bottomrule
			&FLOPs(G) &  \#Params(M)  &EER(\%) &minDCF  \\
			\hline			
			\hline
			DS-TDNN-B                                     & 2.1{\tiny  $_{\,  0\%}$}     
			& 13.2{\tiny  $_{\,  0\%}$}
			& 0.78{\tiny $_{\, 0\%}$}    
			& 0.092{\tiny $_{\, 0\%}$} \\
			\hline	
			MSA$^{\dagger}$ $\rightarrow$ DGF layer       		      & 2.6{\tiny  $_{\,  \textcolor{red}{24\%\uparrow}}$}   
			& 13.7{\tiny $_{\, 4\%\uparrow}$}
			& 0.79{\tiny $_{\, 1\%\uparrow}$}    
			& 0.097{\tiny$_{\, 5\%\uparrow}$}                          \\
			LSTM$\rightarrow$ DGF layer       			  & 3.4{\tiny  $_{\,  \textcolor{red}{62\%\uparrow}}$}   
			& 17.6{\tiny $_{\, \textcolor{red}{44\%\uparrow}}$}
			& 0.84{\tiny $_{\, 8\%\uparrow}$}    
			& 0.108{\tiny$_{\, 17\%\uparrow}$}                          \\
			Bi-LSTM$\rightarrow$ DGF layer       		  & 4.9{\tiny  $_{\,  \textcolor{red}{133\%\uparrow}}$}   
			& 24.7{\tiny $_{\, \textcolor{red}{87\%\uparrow}}$}
			& 0.81{\tiny $_{\, 4\%\uparrow}$}    
			& 0.103{\tiny$_{\, 12\%\uparrow}$}                          \\
			\toprule 
	\end{tabular}} 
	\begin{tablenotes}
		\footnotesize
		\item[] $\dagger$ MSA has 4 heads.
	\end{tablenotes}
\end{table}
\begin{figure}[!t]
	\centering
	\hspace{-2mm}
	\subfloat{
		\includegraphics[width=0.48\linewidth]{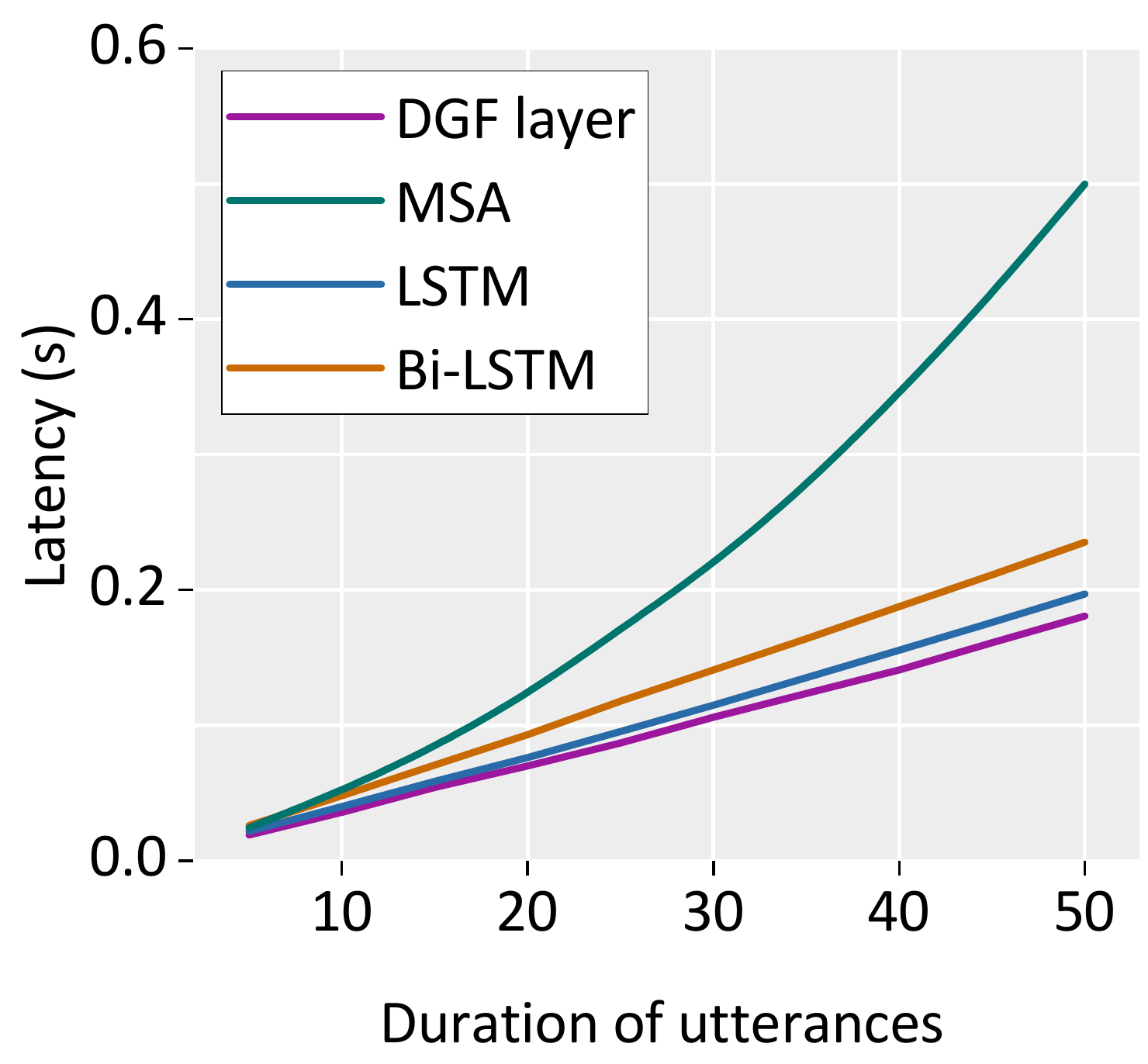}
	}
	\subfloat{
		\includegraphics[width=0.48\linewidth]{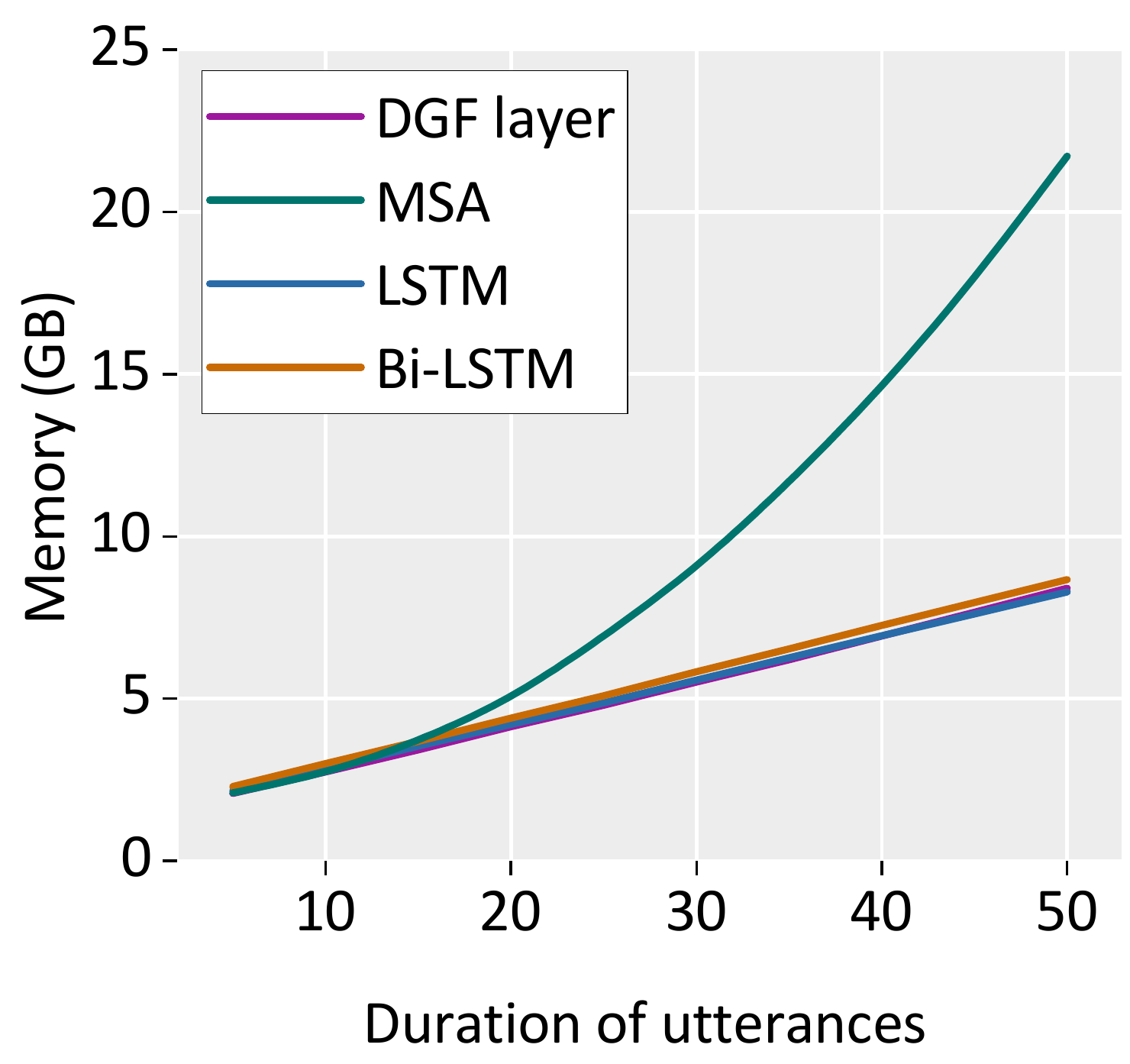}
	}
	\caption{Latency and memory usage for variants of DS-TDNN in which the DGF layer is replaced by various global module designs, measured under a single NVIDIA A5000 GPU and a batch size of 16.}
	\label{fig 6}
\end{figure}

\subsection{Ablation study on the global module design} 

From Fig. \ref{fig 6}, it is seen that the DGF layer has the lowest complexity compared with the other three global module designs. In particular, the multi-head self-attention scheme is most exhaustive in terms of inferring time. Additionally, the DGF layer achieves the best results in EER and minDCF, as reported in Table \ref{Table 5}.
Notably, MSA is comparable to DGF in terms of performance, but it has an obviously higher computation cost. Hence, DGF can be an efficient alternative to MSA when computational cost is a major concern. It is also noted in the table that the LSTM can only achieve suboptimal performance for the given task, which may be accounted by the undirected property of speaker information. This problem is somewhat alleviated in Bi-LSTM by preserving bidirectional information flow.

\begin{table}[t]
	\caption{Ablation study on the macro design. \protect\sethlcolor{mygray}\hl{GRAY} Denotes the DS-TDNN-B}
	\centering
	\label{Table 6}
	\renewcommand\arraystretch{1.5}
	\setlength{\tabcolsep}{1.2mm}{
		\begin{tabular}{l| l l l l}
			\bottomrule
			&FLOPs (G) &  \#Params (M)  &EER (\%) &minDCF  \\			
			\hline
			\hline
			Blocks $= [2, 0]\times 3$ & 2.2{\tiny  $_{\,   \textcolor{red}{35\%\downarrow}}$}   
			& 12.2{\tiny $_{\,  \textcolor{red}{21\%\downarrow}}$}
			& 0.95{\tiny $_{\, 8\%\uparrow}$}      & 0.127{\tiny $_{\, 12\%\uparrow}$}  \\
			Blocks $= [0, 2]\times 3$ & 2.1{\tiny  $_{\,  5\%\downarrow}$}   & 14.3{\tiny  $_{\,  17\%\uparrow}$}
			& 1.07{\tiny $_{\, 13\%\uparrow}$}     & 0.135{\tiny $_{\, 6\%\uparrow}$}   \\
			Blocks $= [1, 1]\times 3$ & 2.1{\tiny  $_{\,  2\%\downarrow}$}   & 13.2{\tiny  $_{\,  8\%\downarrow}$}
			& 0.84{\tiny $_{\,  \textcolor{red}{21\%\downarrow}}$}   
			& 0.098{\tiny$_{\, \textcolor{red}{27\%\downarrow}}$}                       \\
			\rowcolor{mygray}						      
			+  Info exchange          & 2.1{\tiny  $_{\,  0\%\uparrow}$}     & 13.2{\tiny  $_{\,  0\%\uparrow}$}
			& 0.78{\tiny $_{\, 7\%\downarrow}$}    & 0.092{\tiny $_{\, 6\%\downarrow}$} \\
			- MFA       			  & 1.5{\tiny  $_{\,  \textcolor{red}{29\%\downarrow}}$}   
			& 10.0{\tiny $_{\, \textcolor{red}{24\%\downarrow}}$}
			& 0.92{\tiny $_{\, 18\%\uparrow}$}    
			& 0.119{\tiny$_{\,\textcolor{red}{29\%\uparrow}}$}                          \\		
			\hline
	\end{tabular}}
\end{table}
\subsection{Ablation study on the macro design} 
In this part, we study the impact of every macro design, including the dual-stream framework, the information exchange between the local and global branch and the feature aggregation strategy for different-scale information. The detailed results are presented in Table \ref{Table 6}. Notably, the dual-branch framework surpasses its single-branch counterparts with comparable computational overhead, while utilizing local branch shows better performance than utilizing global branch. Besides, the element-wise summation at the end of each block further reduces the EER and the minDCF without increasing complexity. Finally, the MFA proposed in ECAPA also benefits DS-TDNN, even though it obviously increases the complexity.

\begin{table}[t]
	\caption{Ablation Study. \protect\sethlcolor{mygray}\hl{GRAY} Denotes the DS-TDNN-B}
	\centering
	\label{Table 7}
	\renewcommand\arraystretch{1.5}
	\setlength{\tabcolsep}{1.8mm}{
		\begin{tabular}{cc| c c}
			\bottomrule
			Experts $K$  &Sparse ratio &EER (\%) &minDCF  \\
			\hline
			\hline
			$[4,\ 8,\ 8]$& w/o SR             & 0.82  & 0.098\\
			$[4,\ 8,\ 8]$&$[0.3,\ 0.3,\ 0.3]$ & 0.88  & 0.106\\
			$[4,\ 8,\ 8]$&$[0.3,\ 0.3,\ 0.1]$ & 0.83  & 0.096\\
			\rowcolor{mygray}	
			$[4,\ 8,\ 8]$&$[0.3,\ 0.1,\ 0.1]$ & 0.78  & 0.092\\
			$[4,\ 8,\ 8]$&$[0.2,\ 0.1,\ 0.1]$ & 0.79  & 0.093\\
			\hline
			\hline
			-            &w/o SR              & 0.84 & 0.105\\
			-            &$[0.3,\ 0.3,\ 0.3]$ & 0.91 & 0.117\\
			-            &$[0.3,\ 0.3,\ 0.1]$ & 0.86 & 0.109\\
			-            &$[0.3,\ 0.1,\ 0.1]$ & 0.82 & 0.102\\
			-            &$[0.2,\ 0.1,\ 0.1]$ & 0.82 & 0.104\\
			\toprule 
	\end{tabular}}
\end{table}

\subsection{Ablation study on the dynamic filters}
 In Table \ref{Table 7}, the impact of dynamic filters and sparse regularization is investigated. From the table, it is seen that the system employing dynamic filters outperforms the one using static filters. Besides, the proposed Sparse Regularization (SR)  enhances the performance of both dynamic global-aware filters and static global-aware filters. Moreover, a higher sparse ratio does not always guarantee better performance, and an excessively high sparse ratio even lead to an obvious decline in performance. The experimental results also show that the shallow layers generally require a higher sparsity ratio since inactive filters appear more frequently in shallow layers.

\begin{table}[t]
	\begin{center}
		\caption{Ablation Study on the depth and width of the network. \protect\sethlcolor{mygray}\hl{GRAY} Denotes the DS-TDNN-B}
		\label{Table 8}
		\renewcommand\arraystretch{1.5}
		\setlength{\tabcolsep}{1mm}{
			\begin{tabular}{l|ccccccc}
				\bottomrule
				&Layer & Channel &FLOPs (G)&\#Params (M) &EER (\%)   &minDCF \\
				\hline
				\hline
				\rowcolor{mygray}	
				Basic             & $3$   & $1024$  & 2.1    & 13.2       & 0.78     & 0.092 \\	
				\hline
				Width $\uparrow$  & $3$   & $1280$  & 2.8    & 17.1       & 0.70     & 0.088 \\
				Width $\uparrow$  & $3$   & $1536$  & 3.2    & 20.5       & 0.64     & 0.082 \\
				\hline
				Depth $\uparrow$ & $4$   & $1024$  & 2.9    & 17.0       & 0.74     & 0.090 \\
				Depth $\uparrow$ & $5$   & $1024$  & 3.2    & 20.8       & 0.72     & 0.087 \\
				
				\toprule
		\end{tabular}}
	\end{center}
\end{table}
\subsection{Ablation study on the depth and width of the network}
Finally, to reveal the effect of the network's depth and width, various scaling factors are considered in the experiments. Table \ref{Table 8} reveals that DS-TDNN has excellent scalability in both width and depth, while increasing the width of the model is more effective than increasing the depth. Although this seems to be contrary to the conclusion of the recent work \cite{liu2023depth}, it is still reasonable because the DGF layer applied in DS-TDNN is able to provide the global context in the shallow layers, making it unnecessary to enlarge the receptive field by stacking more hidden layers in depth.

\section{Conclusion}

In this paper, we propose a novel global-aware filter (GF) layer to capture long-term context in utterances. The GF layer has global receptive fields while maintaining log-linear complexity. Additionally, we propose dynamic filtering and sparse regularization to enhance the GF layer, which improves its representation and generalization ability. Thereafter, a Dual-Stream Time-Delay Neural Network (DS-TDNN) is built by incorporating the GF layer. The DS-TDNN disentangles the local features and global context, then refines them in a proposed parallel framework with several carefully designed strategies. Experiments on the Voxceleb datasets demonstrate that DS-TDNN achieves a 10\% improvement in EER but with a 28\% and 15\% decline in complexity over ECAPA-TDNN for ASV task using short utterances. Moreover, it outperforms SE-ResNet, AST, and MFA-Conformer with an approximate parameter overhead. Experiments on the SITW datasets reveal that the explicit modeling for global context in DS-TDNN further boosts its performance over ECAPA-DTNN and SE-ResNet when the utterance's duration increases. For extremely long utterances (over 50 seconds), DS-TDNN offers the best trade-off between performance and computational cost, highlighting the advantages of both the GF layer and several designs used in DS-TDNN. This study fully explores the potential of TDNN in speaker verification from the point of both local and global modeling. It might provide some new insights for future network designs of deep speaker embedding or related fields.

\bibliographystyle{IEEEtran}
\bibliography{mybib}

\begin{thebibliography}{10}
\providecommand{\url}[1]{#1}
\csname url@samestyle\endcsname
\providecommand{\newblock}{\relax}
\providecommand{\bibinfo}[2]{#2}
\providecommand{\BIBentrySTDinterwordspacing}{\spaceskip=0pt\relax}
\providecommand{\BIBentryALTinterwordstretchfactor}{4}
\providecommand{\BIBentryALTinterwordspacing}{\spaceskip=\fontdimen2\font plus
\BIBentryALTinterwordstretchfactor\fontdimen3\font minus
  \fontdimen4\font\relax}
\providecommand{\BIBforeignlanguage}[2]{{%
\expandafter\ifx\csname l@#1\endcsname\relax
\typeout{** WARNING: IEEEtran.bst: No hyphenation pattern has been}%
\typeout{** loaded for the language `#1'. Using the pattern for}%
\typeout{** the default language instead.}%
\else
\language=\csname l@#1\endcsname
\fi
#2}}
\providecommand{\BIBdecl}{\relax}
\BIBdecl

\bibitem{rosenberg1976automatic}
A.~E. Rosenberg, ``Automatic speaker verification: A review,''
  \emph{Proceedings of the IEEE}, vol.~64, no.~4, pp. 475--487, 1976.

\bibitem{broun2002automatic}
C.~C. Broun, X.~Zhang, R.~M. Mersereau, and M.~Clements, ``Automatic
  speechreading with application to speaker verification,'' in \emph{2002 IEEE
  International Conference on Acoustics, Speech, and Signal Processing},
  vol.~1.\hskip 1em plus 0.5em minus 0.4em\relax IEEE, 2002, pp. I--685.

\bibitem{becker2008forensic}
T.~Becker, M.~Jessen, and C.~Grigoras, ``Forensic speaker verification using
  formant features and gaussian mixture models,'' in \emph{Ninth Annual
  Conference of the International Speech Communication Association}, 2008.

\bibitem{reynolds2000speaker}
D.~A. Reynolds, T.~F. Quatieri, and R.~B. Dunn, ``Speaker verification using
  adapted gaussian mixture models,'' \emph{Digital signal processing}, vol.~10,
  no. 1-3, pp. 19--41, 2000.

\bibitem{kenny2007joint}
P.~Kenny, G.~Boulianne, P.~Ouellet, and P.~Dumouchel, ``Joint factor analysis
  versus eigenchannels in speaker recognition,'' \emph{IEEE Transactions on
  Audio, Speech, and Language Processing}, vol.~15, no.~4, pp. 1435--1447,
  2007.

\bibitem{dehak2010front}
N.~Dehak, P.~J. Kenny, R.~Dehak, P.~Dumouchel, and P.~Ouellet, ``Front-end
  factor analysis for speaker verification,'' \emph{IEEE Transactions on Audio,
  Speech, and Language Processing}, vol.~19, no.~4, pp. 788--798, 2010.

\bibitem{lei2014novel}
Y.~Lei, N.~Scheffer, L.~Ferrer, and M.~McLaren, ``A novel scheme for speaker
  recognition using a phonetically-aware deep neural network,'' in \emph{2014
  IEEE international conference on acoustics, speech and signal processing
  (ICASSP)}.\hskip 1em plus 0.5em minus 0.4em\relax IEEE, 2014, pp. 1695--1699.

\bibitem{variani2014deep}
E.~Variani, X.~Lei, E.~McDermott, I.~L. Moreno, and J.~Gonzalez-Dominguez,
  ``Deep neural networks for small footprint text-dependent speaker
  verification,'' in \emph{2014 IEEE international conference on acoustics,
  speech and signal processing (ICASSP)}.\hskip 1em plus 0.5em minus
  0.4em\relax IEEE, 2014, pp. 4052--4056.

\bibitem{snyder2017deep}
D.~Snyder, D.~Garcia-Romero, D.~Povey, and S.~Khudanpur, ``Deep neural network
  embeddings for text-independent speaker verification.'' in
  \emph{Interspeech}, vol. 2017, 2017, pp. 999--1003.

\bibitem{waibel1989phoneme}
A.~Waibel, T.~Hanazawa, G.~Hinton, K.~Shikano, and K.~J. Lang, ``Phoneme
  recognition using time-delay neural networks,'' \emph{IEEE transactions on
  acoustics, speech, and signal processing}, vol.~37, no.~3, pp. 328--339,
  1989.

\bibitem{novoselov2018deep}
S.~Novoselov, A.~Shulipa, I.~Kremnev, A.~Kozlov, and V.~Shchemelinin, ``On deep
  speaker embeddings for text-independent speaker recognition,'' \emph{arXiv
  preprint arXiv:1804.10080}, 2018.

\bibitem{he2016deep}
K.~He, X.~Zhang, S.~Ren, and J.~Sun, ``Deep residual learning for image
  recognition,'' in \emph{Proceedings of the IEEE conference on computer vision
  and pattern recognition}, 2016, pp. 770--778.

\bibitem{snyder2019speaker}
D.~Snyder, D.~Garcia-Romero, G.~Sell, A.~McCree, D.~Povey, and S.~Khudanpur,
  ``Speaker recognition for multi-speaker conversations using x-vectors,'' in
  \emph{ICASSP 2019-2019 IEEE International conference on acoustics, speech and
  signal processing (ICASSP)}.\hskip 1em plus 0.5em minus 0.4em\relax IEEE,
  2019, pp. 5796--5800.

\bibitem{huang2019deeper}
X.~Huang, W.~Zhang, X.~Xu, R.~Yin, and D.~Chen, ``Deeper time delay neural
  networks for effective acoustic modelling,'' in \emph{Journal of Physics:
  Conference Series}, vol. 1229, no.~1.\hskip 1em plus 0.5em minus 0.4em\relax
  IOP Publishing, 2019, p. 012076.

\bibitem{srivastava2014dropout}
N.~Srivastava, G.~Hinton, A.~Krizhevsky, I.~Sutskever, and R.~Salakhutdinov,
  ``Dropout: a simple way to prevent neural networks from overfitting,''
  \emph{The journal of machine learning research}, vol.~15, no.~1, pp.
  1929--1958, 2014.

\bibitem{yu2020densely}
Y.-Q. Yu and W.-J. Li, ``Densely connected time delay neural network for
  speaker verification.'' in \emph{INTERSPEECH}, 2020, pp. 921--925.

\bibitem{zhang2020aret}
R.~Zhang, J.~Wei, W.~Lu, L.~Wang, M.~Liu, L.~Zhang, J.~Jin, and J.~Xu, ``Aret:
  Aggregated residual extended time-delay neural networks for speaker
  verification.'' in \emph{INTERSPEECH}, 2020, pp. 946--950.

\bibitem{desplanques2020ecapa}
B.~Desplanques, J.~Thienpondt, and K.~Demuynck, ``Ecapa-tdnn: Emphasized
  channel attention, propagation and aggregation in tdnn based speaker
  verification,'' \emph{arXiv preprint arXiv:2005.07143}, 2020.

\bibitem{gao2019res2net}
S.-H. Gao, M.-M. Cheng, K.~Zhao, X.-Y. Zhang, M.-H. Yang, and P.~Torr,
  ``Res2net: A new multi-scale backbone architecture,'' \emph{IEEE transactions
  on pattern analysis and machine intelligence}, vol.~43, no.~2, pp. 652--662,
  2019.

\bibitem{9688119}
Z.~Li, C.~Fang, R.~Xiao, W.~Wang, and Y.~Yan, ``Si-net: Multi-scale
  context-aware convolutional block for speaker verification,'' in \emph{2021
  IEEE Automatic Speech Recognition and Understanding Workshop (ASRU)}, 2021,
  pp. 220--227.

\bibitem{gu2021dynamic}
B.~Gu and W.~Guo, ``Dynamic convolution with global-local information for
  session-invariant speaker representation learning,'' \emph{IEEE Signal
  Processing Letters}, vol.~29, pp. 404--408, 2021.

\bibitem{mun2022selective}
S.~H. Mun, J.-w. Jung, and N.~S. Kim, ``Selective kernel attention for robust
  speaker verification,'' \emph{arXiv preprint arXiv:2204.01005}, 2022.

\bibitem{chen2019speaker}
C.-P. Chen, S.-Y. Zhang, C.-T. Yeh, J.-C. Wang, T.~Wang, and C.-L. Huang,
  ``Speaker characterization using tdnn-lstm based speaker embedding,'' in
  \emph{ICASSP 2019-2019 IEEE International Conference on Acoustics, Speech and
  Signal Processing (ICASSP)}.\hskip 1em plus 0.5em minus 0.4em\relax IEEE,
  2019, pp. 6211--6215.

\bibitem{jiang2019effective}
Y.~Jiang, Y.~Song, I.~McLoughlin, Z.~Gao, and L.-R. Dai, ``An effective deep
  embedding learning architecture for speaker verification.'' in
  \emph{INTERSPEECH}, 2019, pp. 4040--4044.

\bibitem{sak2014long}
H.~Sak, A.~Senior, and F.~Beaufays, ``Long short-term memory based recurrent
  neural network architectures for large vocabulary speech recognition,''
  \emph{arXiv preprint arXiv:1402.1128}, 2014.

\bibitem{liu2019speaker}
H.~Liu and L.~Zhao, ``A speaker verification method based on tdnn--lstmp,''
  \emph{Circuits, Systems, and Signal Processing}, vol.~38, pp. 4840--4854,
  2019.

\bibitem{tu2019towards}
M.~Tu, Y.~Tang, J.~Huang, X.~He, and B.~Zhou, ``Towards adversarial learning of
  speaker-invariant representation for speech emotion recognition,''
  \emph{arXiv preprint arXiv:1903.09606}, 2019.

\bibitem{lin2019lstm}
Q.~Lin, R.~Yin, M.~Li, H.~Bredin, and C.~Barras, ``Lstm based similarity
  measurement with spectral clustering for speaker diarization,'' \emph{arXiv
  preprint arXiv:1907.10393}, 2019.

\bibitem{huang2020speaker}
C.-L. Huang, ``Speaker characterization using tdnn, tdnn-lstm,
  tdnn-lstm-attention based speaker embeddings for nist sre 2019.'' in
  \emph{Odyssey}, 2020, pp. 423--427.

\bibitem{li2020falcon}
S.~Li, K.~Xue, B.~Zhu, C.~Ding, X.~Gao, D.~Wei, and T.~Wan, ``Falcon: A fourier
  transform based approach for fast and secure convolutional neural network
  predictions,'' in \emph{Proceedings of the IEEE/CVF Conference on Computer
  Vision and Pattern Recognition}, 2020, pp. 8705--8714.

\bibitem{chi2020fast}
L.~Chi, B.~Jiang, and Y.~Mu, ``Fast fourier convolution,'' \emph{Advances in
  Neural Information Processing Systems}, vol.~33, pp. 4479--4488, 2020.

\bibitem{lifourier}
Z.~Li, N.~B. Kovachki, K.~Azizzadenesheli, K.~Bhattacharya, A.~Stuart,
  A.~Anandkumar \emph{et~al.}, ``Fourier neural operator for parametric partial
  differential equations,'' in \emph{International Conference on Learning
  Representations}.

\bibitem{yang2019condconv}
B.~Yang, G.~Bender, Q.~V. Le, and J.~Ngiam, ``Condconv: Conditionally
  parameterized convolutions for efficient inference,'' \emph{Advances in
  Neural Information Processing Systems}, vol.~32, 2019.

\bibitem{chen2020dynamic}
Y.~Chen, X.~Dai, M.~Liu, D.~Chen, L.~Yuan, and Z.~Liu, ``Dynamic convolution:
  Attention over convolution kernels,'' in \emph{Proceedings of the IEEE/CVF
  conference on computer vision and pattern recognition}, 2020, pp.
  11\,030--11\,039.

\bibitem{li2022omni}
C.~Li, A.~Zhou, and A.~Yao, ``Omni-dimensional dynamic convolution,''
  \emph{arXiv preprint arXiv:2209.07947}, 2022.

\bibitem{liu2022mfa}
T.~Liu, R.~K. Das, K.~A. Lee, and H.~Li, ``Mfa: Tdnn with multi-scale
  frequency-channel attention for text-independent speaker verification with
  short utterances,'' in \emph{ICASSP 2022-2022 IEEE International Conference
  on Acoustics, Speech and Signal Processing (ICASSP)}.\hskip 1em plus 0.5em
  minus 0.4em\relax IEEE, 2022, pp. 7517--7521.

\bibitem{zhang2022mfa}
Y.~Zhang, Z.~Lv, H.~Wu, S.~Zhang, P.~Hu, Z.~Wu, H.-y. Lee, and H.~Meng,
  ``Mfa-conformer: Multi-scale feature aggregation conformer for automatic
  speaker verification,'' \emph{arXiv preprint arXiv:2203.15249}, 2022.

\bibitem{okabe2018attentive}
K.~Okabe, T.~Koshinaka, and K.~Shinoda, ``Attentive statistics pooling for deep
  speaker embedding,'' \emph{arXiv preprint arXiv:1803.10963}, 2018.

\bibitem{hu2018squeeze}
J.~Hu, L.~Shen, and G.~Sun, ``Squeeze-and-excitation networks,'' in
  \emph{Proceedings of the IEEE conference on computer vision and pattern
  recognition}, 2018, pp. 7132--7141.

\bibitem{nagrani2017voxceleb}
A.~Nagrani, J.~S. Chung, and A.~Zisserman, ``Voxceleb: a large-scale speaker
  identification dataset,'' \emph{arXiv preprint arXiv:1706.08612}, 2017.

\bibitem{chung2018voxceleb2}
J.~S. Chung, A.~Nagrani, and A.~Zisserman, ``Voxceleb2: Deep speaker
  recognition,'' \emph{arXiv preprint arXiv:1806.05622}, 2018.

\bibitem{mclaren2016speakers}
M.~McLaren, L.~Ferrer, D.~Castan, and A.~Lawson, ``The speakers in the wild
  (sitw) speaker recognition database.'' in \emph{Interspeech}, 2016, pp.
  818--822.

\bibitem{snyder2015musan}
D.~Snyder, G.~Chen, and D.~Povey, ``Musan: A music, speech, and noise corpus,''
  \emph{arXiv preprint arXiv:1510.08484}, 2015.

\bibitem{ko2017study}
T.~Ko, V.~Peddinti, D.~Povey, M.~L. Seltzer, and S.~Khudanpur, ``A study on
  data augmentation of reverberant speech for robust speech recognition,'' in
  \emph{2017 IEEE International Conference on Acoustics, Speech and Signal
  Processing (ICASSP)}.\hskip 1em plus 0.5em minus 0.4em\relax IEEE, 2017, pp.
  5220--5224.

\bibitem{park2019specaugment}
D.~S. Park, W.~Chan, Y.~Zhang, C.-C. Chiu, B.~Zoph, E.~D. Cubuk, and Q.~V. Le,
  ``Specaugment: A simple data augmentation method for automatic speech
  recognition,'' \emph{arXiv preprint arXiv:1904.08779}, 2019.

\bibitem{chung2020defence}
J.~S. Chung, J.~Huh, S.~Mun, M.~Lee, H.~S. Heo, S.~Choe, C.~Ham, S.~Jung, B.-J.
  Lee, and I.~Han, ``In defence of metric learning for speaker recognition,''
  \emph{arXiv preprint arXiv:2003.11982}, 2020.

\bibitem{zhao2021speakin}
M.~Zhao, Y.~Ma, M.~Liu, and M.~Xu, ``The speakin system for voxceleb speaker
  recognition challange 2021,'' \emph{arXiv preprint arXiv:2109.01989}, 2021.

\bibitem{shim2022graph}
H.-j. Shim, J.~Heo, J.-h. Park, G.-h. Lee, and H.-J. Yu, ``Graph attentive
  feature aggregation for text-independent speaker verification,'' in
  \emph{ICASSP 2022-2022 IEEE International Conference on Acoustics, Speech and
  Signal Processing (ICASSP)}.\hskip 1em plus 0.5em minus 0.4em\relax IEEE,
  2022, pp. 7972--7976.

\bibitem{gong2021ast}
Y.~Gong, Y.-A. Chung, and J.~Glass, ``Ast: Audio spectrogram transformer,''
  \emph{Interspeech}, 2021.

\bibitem{thienpondt2021idlab}
J.~Thienpondt, B.~Desplanques, and K.~Demuynck, ``The idlab voxsrc-20
  submission: Large margin fine-tuning and quality-aware score calibration in
  dnn based speaker verification,'' in \emph{ICASSP 2021-2021 IEEE
  International Conference on Acoustics, Speech and Signal Processing
  (ICASSP)}.\hskip 1em plus 0.5em minus 0.4em\relax IEEE, 2021, pp. 5814--5818.

\bibitem{deng2019arcface}
J.~Deng, J.~Guo, N.~Xue, and S.~Zafeiriou, ``Arcface: Additive angular margin
  loss for deep face recognition,'' in \emph{Proceedings of the IEEE/CVF
  conference on computer vision and pattern recognition}, 2019, pp. 4690--4699.

\bibitem{xiang2019margin}
X.~Xiang, S.~Wang, H.~Huang, Y.~Qian, and K.~Yu, ``Margin matters: Towards more
  discriminative deep neural network embeddings for speaker recognition,'' in
  \emph{2019 Asia-Pacific Signal and Information Processing Association Annual
  Summit and Conference (APSIPA ASC)}.\hskip 1em plus 0.5em minus 0.4em\relax
  IEEE, 2019, pp. 1652--1656.

\bibitem{kingma2014adam}
D.~P. Kingma and J.~Ba, ``Adam: A method for stochastic optimization,''
  \emph{arXiv preprint arXiv:1412.6980}, 2014.

\bibitem{cumani2011comparison}
S.~Cumani, P.~D. Batzu, D.~Colibro, C.~Vair, P.~Laface, and V.~Vasilakakis,
  ``Comparison of speaker recognition approaches for real applications.'' in
  \emph{INTERSPEECH}, 2011, pp. 2365--2368.

\bibitem{parkvision}
N.~Park and S.~Kim, ``How do vision transformers work?'' in \emph{International
  Conference on Learning Representations}.

\bibitem{cordonnierrelationship}
J.-B. Cordonnier, A.~Loukas, and M.~Jaggi, ``On the relationship between
  self-attention and convolutional layers,'' in \emph{International Conference
  on Learning Representations}.

\bibitem{siinception}
C.~Si, W.~Yu, P.~Zhou, Y.~Zhou, X.~Wang, and Y.~Shuicheng, ``Inception
  transformer,'' in \emph{Advances in Neural Information Processing Systems}.

\end{thebibliography}

\end{document}